\documentclass[12pt,a4paper]{article}

\usepackage[utf8]{inputenc}
\usepackage[T1]{fontenc}
\usepackage{libertine}
\usepackage{amsmath,amssymb}
\usepackage{graphicx}
\usepackage{booktabs}
\usepackage{multirow}
\usepackage[hypertexnames=false,hidelinks]{hyperref}
\usepackage{url}
\usepackage{listings}
\usepackage{xcolor}
\usepackage{float}
\usepackage{microtype}
\usepackage[margin=2.5cm]{geometry}
\usepackage{tikz}
\definecolor{inlinecode}{RGB}{0,100,100}
\newcommand{\code}[1]{{\texttt{\small\color{inlinecode}#1}}}

\definecolor{codebg}{gray}{0.96}
\definecolor{tomlbg}{RGB}{255,250,240}
\definecolor{rustbg}{RGB}{240,245,255}
\definecolor{shellbg}{gray}{0.94}
\definecolor{codegreen}{rgb}{0.0,0.5,0.0}
\definecolor{codegray}{rgb}{0.5,0.5,0.5}
\definecolor{codeblue}{rgb}{0.0,0.0,0.7}
\definecolor{rustorange}{rgb}{0.7,0.3,0.0}
\definecolor{shellprompt}{rgb}{0.3,0.3,0.3}

\lstdefinestyle{toml}{
  backgroundcolor=\color{tomlbg},
  basicstyle=\ttfamily\small,
  breaklines=true,
  keywordstyle=\color{codeblue}\bfseries,
  commentstyle=\color{codegray}\itshape,
  stringstyle=\color{codegreen},
  frame=l,
  framerule=1.5pt,
  rulecolor=\color{codeblue},
  xleftmargin=6pt,
  xrightmargin=2pt,
}

\lstdefinestyle{rust}{
  backgroundcolor=\color{rustbg},
  basicstyle=\ttfamily\small,
  breaklines=true,
  keywordstyle=\color{rustorange}\bfseries,
  commentstyle=\color{codegray}\itshape,
  stringstyle=\color{codegreen},
  frame=l,
  framerule=1.5pt,
  rulecolor=\color{rustorange},
  xleftmargin=6pt,
  xrightmargin=2pt,
}

\definecolor{pythonbg}{RGB}{245,252,245}
\lstdefinestyle{python}{
  language=Python,
  backgroundcolor=\color{pythonbg},
  basicstyle=\ttfamily\small,
  breaklines=true,
  keywordstyle=\color{codeblue}\bfseries,
  commentstyle=\color{codegray}\itshape,
  stringstyle=\color{codegreen},
  frame=l,
  framerule=1.5pt,
  rulecolor=\color{codegreen},
  xleftmargin=6pt,
  xrightmargin=2pt,
}

\lstdefinestyle{shell}{
  backgroundcolor=\color{shellbg},
  basicstyle=\ttfamily\small,
  breaklines=true,
  frame=l,
  framerule=1.5pt,
  rulecolor=\color{shellprompt},
  xleftmargin=6pt,
  xrightmargin=2pt,
}

\newcommand{\metricSLOC}{64{,}596}
\newcommand{\metricTests}{1{,}928}
\newcommand{\metricCrates}{25}
\newcommand{\metricDocFiles}{109}

\newcommand{\metricDocTests}{63}           

\title{OxyMake: A Content-Addressed Workflow Engine}

\author{Emmanuel Sérié\\
\small Centre de Mathématiques Appliquées (CMAP), CNRS,\\
\small École Polytechnique, Institut Polytechnique de Paris,\\
\small 91128 Palaiseau Cedex, France}

\date{\today}

\begin{document}

\maketitle

\begin{abstract}
A workflow engine's job is to decide which results survive a change to
code, data, or machine. Almost every engine answers with information
local to where the work happened---a file timestamp, or a provenance
record written beside one copy of the project---so the answer is
worthless anywhere else: the next machine, the next collaborator, the
next checkout recomputes what is already correct. Engines that decide by
content instead often buy the property with infrastructure: a
background service, or a store outside the project.

OxyMake is built on the position that the validity of a result should be
a property of the work, not of the place it ran, and that this need not
cost an operational system. A result's identity is a pure function of
what its rule declares: command text, inputs, parameters, environment.
On a given platform and version of the tool, the same declaration yields
the same identity in every copy of a project whose inputs lie inside it,
decided by a single Rust binary with nothing running behind it. Identity is not reuse---results
must still travel---but it is what makes reuse decidable at all.

We describe the design, measure repeated execution, and model-check
three specifications of the state layer over a bounded space of states.
The guarantees reach exactly as far as the workflow declares, and no
further.
\end{abstract}

\section{Introduction}
\label{sec:introduction}

The validity of a result should be a property of the work, not of the
place it ran. Most engines decide otherwise. Asked which declared
outputs still have a valid derivation, they answer from information
local to one working tree---a file timestamp, or a provenance record
written beside one copy of the project---so the answer does not survive
the move to a second machine, a fresh checkout, or a collaborator, each
of which starts with nothing and recomputes what is already correct.
OxyMake\footnote{Source code and project site:
\url{https://github.com/noogram/oxymake}
(\url{https://oxymake.noogram.dev}). Repository paths cited in this
paper refer to tag \code{v0.1.0}. The head-to-head measurements of
\S\ref{sec:perf-bench}--\S\ref{sec:git-checkout-bench} were taken at
commit \code{03864f8}, not at that tag; the build is stated in
\S\ref{sec:evaluation}.}
takes the opposite position, and takes it with no operational system
behind it. \code{ox run} ensures the declared
outputs exist, executing
only the jobs whose outputs are missing or out of date. The operation is
convergent: repeating it performs only the work that remains, and on an
up-to-date tree it does nothing. Five commands share this model and a
common filter surface---\code{ox run} (converge), \code{ox cancel}
(abort), \code{ox invalidate} (reset), \code{ox plan} (preview), and
\code{ox status} (observe) (\S\ref{sec:idempotent}).

\paragraph{Single-binary execution of a declarative file.}
Requiring no operational system is a design constraint rather than a
packaging property: it is what lets the key be recomputed in a checkout
nobody provisioned.
The workflow is a declarative TOML file; a three-line rule with
\code{input}, \code{output}, and \code{shell} is a valid workflow. It is
executed by a single 17.4\,MB Rust binary that bundles no interpreter
and needs no daemon; \code{ox --help} returns in 2.2\,ms
(\S\ref{sec:startup-bench}, \S\ref{sec:binary-bench}). The same file is
accepted sequentially, in parallel (\code{-j N}), and---by changing one
flag---by the Slurm and Ray backends, neither of which has a supported
deployment configuration in v0.1.0 (\S\ref{sec:distributed-related}).
The evaluation in this paper is single-machine; no distributed
performance is measured (\S\ref{sec:evaluation}).

\paragraph{A machine-readable command interface.}
Every command accepts \code{--json}; when it is given, output is
newline-delimited JSON (NDJSON). Gate approval is a command
(\code{ox gate approve}) rather than a terminal prompt, so a script or an
AI agent drives the same operations a human does; in v0.1.0, however, the
scheduler does not consult gate state, so a declared gate is recorded and
managed but does not block execution (\S\ref{sec:agent-api},
\S\ref{sec:limitations}). An MCP server (\code{ox serve --mcp}) exposes an
inspection subset of the engine over the Model Context Protocol: eight
tools, seven read-only plus \code{ox\_clean}, which removes cached
artifacts and state. This is an
agent-compatible interface; we do not measure agent task success and
make no claim that OxyMake improves it.

\paragraph{A content-derived rebuild decision.}
Each job's cache key (Eq.~\ref{eq:cache-key}) is a BLAKE3 hash of the
rule source, the declared input contents, the parameters, the
environment specification, the shell, and the platform; it records no
state private to the working tree the job ran in. When that declared
specification is unchanged and its inputs lie inside the workflow root,
the \emph{computation key} is therefore identical in any checkout on
the same platform and under the same key-format version ($v$ in
Eq.~\ref{eq:cache-key}).
Snakemake~7 instead records rebuild provenance per working
tree~\cite{kosterSnakemakeScalableBioinformatics2012,molderSustainableDataAnalysis2021};
bound to one tree, that record does not travel, so a fresh clone starts
without it even when equivalent outputs have been restored from
elsewhere---a consequence of the record's scope, not a scenario we
benchmark. What a shared key buys is decidability rather than reuse.
Whether two checkouts take the same \emph{rebuild decision} depends
further on their live reuse state---the outputs materialised on disk,
the selected validation policy, and, under the \code{mtime+hash} default
and \code{hash}, the local key-to-output index---which \S\ref{sec:content-addressable} treats, along with what
does and does not travel between checkouts in v0.1.0. What the shipped
system establishes across checkouts is key identity.

\paragraph{Scope of the content-key claim.}
The key is sound over declared inputs only. OxyMake does not sandbox
rule execution, so an undeclared read---a helper script, a tool on
\code{\$PATH}, an environment variable outside the declared
specification---never enters the key and can produce a stale reuse
(\S\ref{sec:undeclared-inputs}). Output verification depth is a policy:
the default \code{mtime+hash} re-hashes a file when its size or
timestamp changes, \code{mtime} trusts metadata for parity with existing
runners, and \code{hash} re-reads every byte
(\S\ref{sec:content-addressable}). Timestamps serve here as an
experimental control: our benchmark bumps a shared input's
mtime without changing a byte---what a \code{git checkout} does to a
working tree---and counts re-runs. Two of the three outcomes are
measured: Snakemake~7.32.4 re-runs zero jobs, because the recorded
checksum of the perturbed input still matches
(\S\ref{sec:git-checkout-bench}), and a pure-mtime policy re-runs the
full downstream radius. The third follows from the policy definition
and was not measured as a separate condition: under the
\code{mtime+hash} default a
changed timestamp triggers a re-hash, the re-hash finds the same bytes,
and no job can be scheduled---zero re-runs
(\S\ref{sec:git-checkout-bench}). The phantom re-run is
real for GNU Make~\cite{feldmanMakeProgram1979} and for any engine's
pure-mtime fast path, and is not a difference from the benchmarked
Snakemake.

\paragraph{Results.}
On a synthetic 10,001-job workflow on one machine, OxyMake resolves the
job graph in 118.3\,ms against Snakemake~7.32.4's 2.72\,s, a 23.0$\times$
difference on graph resolution (\S\ref{sec:perf-bench}). A cold
end-to-end run is 1.47--2.57$\times$ slower; the differential is
consistent with content-key bookkeeping on the cold path, but was not
phase-profiled; a warm no-op re-run is faster
by a factor that depends on the validation policy: 2.52$\times$ under
the shipped \code{mtime+hash} default, 6.95$\times$ under \code{mtime}
and 1.91$\times$ under \code{hash} (both engines at 16-way
parallelism) (\S\ref{sec:e2e-bench}). Because concurrent \code{ox run} processes
create more interleavings than a test suite can sample, three TLA+
specifications~\cite{lamportSpecifyingSystems2002} model-check bounded
safety properties of the concurrent state protocol
(\S\ref{sec:named-invariants}); this is the hazard class for which
Amazon Web Services adopted TLA+ at production
scale~\cite[pp.~66--73]{newcombeHowAmazonWebServices2015}.

\paragraph{Why a new engine.}
No single property above is new. The combination is new among the
systems we compare (Table~\ref{tab:landscape},
\S\ref{sec:related}); we are not aware of a system outside that set
that combines them, but we have not run a systematic survey. We chose
to build the combination in rather than retrofit it. In the file-rule
engines and build systems closest to OxyMake's design point---GNU
Make, Snakemake, Bazel, Buck2, and Nix---the
default rebuild decision is entangled with a state model or a
deployment premise that an adaptation would have to move: GNU Make's
default decision reads live mtimes, per-tree state by construction;
Snakemake's
default decision reads a per-tree provenance record, and its opt-in
between-workflow cache (\code{--cache}) grafts a content-derived key
beside that record as a per-rule side channel rather than replacing
the default (\S\ref{sec:file-based}); Bazel and Buck2 take the build daemon as
their standard operating mode, and Nix the store as an operating
premise.  (Bazel documents a \code{--batch} startup option that runs
the client without a server, but it is deprecated and loses incremental
state~\cite{bazelCommandLine}.) (DVC is the acknowledged
exception: its default \code{dvc repro} decision already reads content
hashes with no daemon, over pipeline stages rather than wildcard file
rules---\S\ref{sec:file-based}.) Making a key that is a pure function
of declared content---holding no per-tree state and requiring no
external infrastructure---the \emph{default identity for cached
computations} would in each case mean moving that boundary. That
identity is not by itself the rebuild decision: as above, the decision
also consults the outputs materialised on disk, the selected validation
policy, and, under the \code{mtime+hash} default and \code{hash}, the
local key-to-output index. We have not attempted
those adaptations; building the engine around the key is a design
choice, not an impossibility claim.

\paragraph{Contributions.}
This paper contributes:

\begin{enumerate}
\item \textbf{Convergent, daemon-free execution.} \code{ox run}
  reconciles declared outputs against the current tree and is safe to
  repeat, under the stated assumption that declared rules are safe to
  re-run (\S\ref{sec:idempotent}). A persistent claim-state protocol
  is implemented in the state layer and model-checked; it is not yet
  the scheduling gate, so overlapping sessions currently re-execute
  shared jobs independently (\S\ref{sec:named-invariants}).
\item \textbf{Single-binary implementation with pluggable executor
  backends.} A
  17.4\,MB Rust binary with a backward-chaining resolver accepts the
  same workflow file unmodified on a laptop, on Slurm, and under Ray.
  The Slurm and Ray backends are implemented but have no supported
  deployment configuration in v0.1.0, because both need the workspace
  visible to the compute nodes while \code{.oxymake/} must be on local
  disk (\S\ref{sec:implementation},
  \S\ref{sec:distributed-related}, \S\ref{sec:limitations}).
\item \textbf{A machine-readable execution interface.} Structured
  NDJSON events let a program or agent consume a run without scraping
  terminal output, and gate decisions are commands rather than prompts
  (recorded but not yet enforced by the scheduler, \S\ref{sec:agent-api}); an MCP
  server exposes an inspection subset of these operations---seven
  read-only tools plus \code{ox\_clean}
  (\S\ref{sec:agent-api}).
\item \textbf{Content-addressed caching by default.} The cache key is a
  BLAKE3 hash of rule source, declared input contents, parameters,
  environment, shell, and platform (Eq.~\ref{eq:cache-key}); the key is
  identical when that entire declared specification is identical.
  Under the shipped \code{mtime+hash} default and under \code{hash},
  output content is checked at reuse time by the validation policy
  against the local key-to-output index; the opt-in \code{mtime} policy
  is a stateless metadata-only lookup that consults neither the key nor
  the index and verifies no content.  Reuse cascades on deletion
  (\S\ref{sec:content-addressable}).
\item \textbf{Three-graph architecture} (RuleGraph $\to$ JobGraph
  $\to$ ExecGraph) separating rule definition, resolved-and-pruned
  plan, and per-run execution trace, following our three-layer reading
  of the workflow forms described by Goble et
  al.~\cite{gobleFAIRComputationalWorkflows2020}
  (\S\ref{sec:three-graphs}).
\item \textbf{A plan of record (\code{ox.lock})}: a content-hashed
  lockfile that re-derives the same job graph from the same TOML, on
  the same platform, under the recorded assumptions
  (\S\ref{sec:content-addressable}, \S\ref{sec:fair-eval}).
\item \textbf{Three TLA+ specifications} totalling 623 lines that
  model-check bounded safety of the concurrent state protocol
  (\S\ref{sec:named-invariants}).
\end{enumerate}

\paragraph{Running example.}
To ground the discussion that follows, Listing~\ref{lst:running-example}
shows a complete Oxymakefile for a three-stage data pipeline.  The file is
valid TOML; no embedded Python or custom DSL is needed.  Key concepts
introduced here---wildcards, named inputs/outputs, \code{expand}, and the
\code{all} pseudo-rule---are defined precisely in
Sections~\ref{sec:design}--\ref{sec:architecture}.

\begin{lstlisting}[style=toml, caption={A complete Oxymakefile: generate
  per-sample data, compute statistics, and merge into a report.},
  label={lst:running-example}]
ox_version = "0.1"

[config]
samples = ["alpha", "beta", "gamma"]

[rule.all]
input = ["results/report.txt"]

[rule.generate]
output = ["data/{sample}.csv"]
shell = """
echo "word,count" > {output}
for w in the {sample} pipeline workflow; do
    echo "$w,$(( $(printf '%s' "{sample}-$w" \
        | cksum | cut -d' ' -f1) % 100 + 1 ))" >> {output}
done
"""

[rule.stats]
input  = { csv = "data/{sample}.csv" }
output = { txt = "results/{sample}_stats.txt" }
shell  = """
echo "# {sample}: $(tail -n+2 {input.csv} | wc -l) rows" \
     > {output.txt}
"""

[rule.report]
input   = ["results/{sample}_stats.txt"]
output  = ["results/report.txt"]
expand  = "product"
shell   = """
echo "=== Pipeline Report ===" > {output}
cat {input} >> {output}
"""
\end{lstlisting}

\noindent The resolver reads the \code{all} rule, traces its input
\code{results/report.txt} back through \code{report} $\to$ \code{stats}
$\to$ \code{generate}, and expands the \code{\{sample\}} wildcard against
the three configured values: three \code{generate} jobs, three
\code{stats} jobs, and one \code{report} job whose input list expands
(\code{expand = "product"}) to the three per-sample stats files---a
seven-job DAG (\code{all} is a pseudo-target, not a job). The word
counts are derived from a checksum of the sample name and word, so
re-running the workflow reproduces the same bytes. Content-addressable
caching ensures that a re-run after editing only \code{rule.stats} skips
the three \code{generate} jobs entirely.  We refer to this example
throughout the paper.

\section{Background and Related Work}
\label{sec:related}

\subsection{File-Based Workflow Systems}
\label{sec:file-based}

OxyMake inherits Snakemake's core paradigm---backward-chaining DAG
resolution with wildcard-driven genericity---and makes different
trade-offs on three axes: a content-addressed cache key in place of
per-tree provenance, a declarative TOML format in place of an embedded
Python DSL, and a native Rust binary in place of a Python runtime. Each
trade-off has a cost, stated where it is measured (\S\ref{sec:evaluation}).
The systems below are grouped by the axis on which each chiefly
differs from OxyMake.

The lineage of file-based workflow systems begins with
Make~\cite{feldmanMakeProgram1979}, which introduced the fundamental
abstraction: rules declare outputs, inputs, and commands; the engine
constructs a directed acyclic graph (DAG) and executes only the steps whose
outputs are missing or out of date. Every system below inherits some
part of Make's model, but the two halves have different fates: the
backward-chaining resolution model is widely adopted, while the mtime
freshness test that Make pairs it with is not. Several of the systems
below adopt the former and reject the latter, and at least
one---CWL---adopts neither.

\paragraph{The rebuild decision.}
Snakemake~\cite{kosterSnakemakeScalableBioinformatics2012} extended Make
with Python-based rule definitions and multi-wildcard pattern matching;
its 2021 article describes it as ``one of the most widely used workflow
management systems in science'' and covers containerized execution,
cloud integration, and module
composition~\cite{molderSustainableDataAnalysis2021}. Snakemake rules
embed Python, so the full job graph is known only after the workflow
file is executed; static analysis and IDE support are correspondingly
less complete than for an inert format, not absent. In the 7.x line its
change detection adds a recorded per-output checksum to the live mtime
comparison rather than replacing it: an input counts as updated only if
it is newer than the oldest output \emph{and} its recorded SHA-256 no
longer matches, and a checksum is recorded only for inputs smaller than
100{,}000 bytes. Below that threshold the checksum decides, and the
decision survives \code{git checkout} mtime churn unmoved (verified at
every bench scale, \S\ref{sec:git-checkout-bench}); above it the mtime
comparison alone decides. The record is bound to a single
working tree and does not validate output content, so it
is not valid on another machine and does not detect on-disk corruption.
Snakemake also ships an opt-in between-workflow cache (\code{--cache},
with per-rule \code{cache: True}) whose entry name is derived from
rule, input, and software-environment content; it is a per-rule side
channel next to the provenance record, and does not replace the
default rebuild decision.

A neighbouring family versions data pipelines rather than executing
file rules: DVC~\cite{dvc} records content hashes of stage inputs and
outputs in \code{dvc.lock}, a file that travels with the repository,
and takes the \code{dvc repro} rebuild decision from those hashes with
no daemon and an in-project cache. It is a Python runtime layered on
Git rather than a wildcard-driven file-rule engine, and its machine
interface does not extend to a structured event stream, so it does not
occupy the combination claimed here
(\S\ref{sec:discussion})---but it shows that a content-hash
rebuild decision without a daemon predates this work.

\paragraph{The specification format.}
The Common Workflow Language
(CWL)~\cite{crusoeMethodsIncludedStandardizing2022} occupies a different
position in this list: it is a vendor-neutral standard for describing
workflows of command-line tools rather than an engine. Steps are wired
by explicit data connections rather than by filename convention, and
no change-detection policy is prescribed at all. Its documents
declare inbound data links explicitly rather than inferring them.
Because the standard fixes what a step
consumes and produces without prescribing how it is scheduled, staged,
or cached, a conforming engine may optimise freely underneath a workflow
its author never edits. \code{cwltool}, the reference implementation, is
scoped for feature-completeness and validation rather than for
portability to every host operating system, and already offers an
optional cache: with \code{--cachedir} it fingerprints the realised
command, content checksums where available (otherwise size and mtime),
container selection, and selected requirements, naming the cache entry
from a digest of that dictionary. It is therefore content-sensitive
where checksums exist, and metadata-sensitive where they do not;
platforms such as Arvados~\cite{arvados} exercise the standard's
latitude further, pairing CWL with content-addressed data
management. CWL is therefore better read as a complementary layer to
this work than as a competitor to it---a description language OxyMake
could consume, rather than a change-detection policy it must
correct.

Galaxy~\cite{afganGalaxyPlatformAccessible2018}
leads with a web-based interface built for biologists; it also exposes
a REST API with language bindings for programmatic use, while
Planemo provides a command-line tool for developing and testing Galaxy
tools and workflows---rather than
the text-first authoring loop this paper assumes.

The Workflow
Description Language (WDL)~\cite{vossWDLCromwell2017} and its reference
engine Cromwell target genomics pipelines with a typed, portable
specification; Cromwell is backend-agnostic, with a pre-enabled Local
backend as its default and pluggable HPC, TES and cloud-batch backends
behind the same interface. One-shot \code{cromwell run} executes a
workflow as a CLI process and exits; a server deployment is required
only for its server-mode operations. Cromwell also offers a call cache,
disabled by default, keyed on the task command, the task inputs (input
files hashed by content, with \code{md5} among the configurable local
filesystem hashing strategies)
and the \code{docker}, \code{continueOnReturnCode} and
\code{failOnStderr} runtime attributes, with a floating image tag
resolved to its immutable digest; that reuse is bound to a persistent
Cromwell database rather than to the workflow declaration. The cost
relative to OxyMake is therefore a separate execution engine, a
more verbose specification
(\S\ref{sec:snakemake-compat}), and a reuse scope tied to a database
deployment.

\paragraph{The runtime and what it presupposes.}
Nextflow~\cite{diTommasoNextflowEnablesReproducible2017} takes a
dataflow-oriented approach with channels connecting processes, with
strong containerization support. Its \code{-resume} rebuild decision is
an automatic per-task hash over the task script, the container, Conda or
Spack specification, the task inputs and the variables the script
references, computed with no daemon; input files enter that hash by
name, size and last-modified timestamp by default, and by content under
the \code{cache 'deep'} directive. The axis that separates it from
OxyMake is therefore the Groovy DSL and the JVM runtime rather than the
absence of a content key.

Pegasus~\cite{deelmanPegasusWorkflowManagement2015} targets workflows
from localhost execution---which its shell planner supports without
HTCondor---to large distributed resources; its distributed modes
introduce execution and data-staging infrastructure beyond OxyMake's
local design point.

\subsection{Build System Theory}
\label{sec:build-theory}

Mokhov et al.~\cite[\S3--4]{mokhovBuildSystemsCarte2018}~\cite[\S4--5]{mokhovBuildSystemsCarte2020}
provide the standard theoretical framework for build systems, decomposing
them along two orthogonal axes: the \emph{scheduler}, which decides the
order in which tasks run (topological: fixed order from the dependency graph;
restarting: re-queues a task when a new dependency is discovered; or
suspending: pauses a running task mid-flight to resolve a newly discovered
dependency), and the \emph{rebuilder}, which decides whether a task's
existing output can be reused (dirty bit: rebuild if any input changed;
verifying traces: rebuild only if recorded input/output hashes no longer
match; constructive traces: select a previously built output whose
inputs match; or deep constructive traces: allow intermediate inputs to
differ as long as final content matches). Their key insight
is that every build system is a composition of a scheduler and a rebuilder,
yielding a two-dimensional design space where existing systems (Make,
Shake~\cite{mitchellShakeBuildSystem2012},
Bazel~\cite{bazel}, Nix) occupy specific cells.

In this taxonomy, OxyMake combines a \textbf{topological scheduler} with a
\textbf{verifying-traces rebuilder} augmented with content-addressing,
with one exception: the deleted-output cascade of
\S\ref{sec:idempotent} propagates staleness unconditionally and is a
dirty-bit path. All
dependencies---including wildcard expansions and scatter/gather
patterns---are resolved statically by a backward-chaining resolver before
the scheduler begins execution. The scheduler then computes a topological
order over the fully materialised DAG and dispatches jobs whose upstream
dependencies are satisfied, with no runtime dependency discovery or
suspend/resume. On the scheduler axis this places OxyMake with Make,
Ninja and Buck; the topological/verifying-traces cell it claims is the
one Mokhov et al.\ assign to Ninja. Bazel~\cite{bazel} is not in that
column: it is classified as restarting, because it supports action-time
input discovery (\emph{discovered inputs}), so its evaluation graph is
not fixed before execution in the way an OxyMake JobGraph is. The
affinity with Bazel is elsewhere---its analysis phase likewise
constructs an action graph before execution, and it likewise caches on
content. OxyMake replaces Bazel's Starlark~\cite{starlark}
configuration layer with declarative TOML and adds
domain-specific features (wildcard expansion, environment management, gates)
for scientific workflows.  Meta's Buck2~\cite{buck2} pushes this design further
with a content-hashed, remote-execution-oriented model built on the
Starlark configuration language; OxyMake shares the content-addressing principle but
replaces the Starlark layer with TOML to preserve static parseability.

The distinction between \emph{Applicative} tasks (statically known
dependencies) and \emph{Monadic} tasks (dependencies discovered at runtime)
clarifies OxyMake's position: its rules are strictly Applicative.  Wildcard
patterns and scatter/gather configurations \emph{appear} dynamic but are
expanded at resolution time---before the job graph is constructed---so the
scheduler never needs to suspend a running task to discover new
dependencies.

Dolstra's Nix thesis~\cite[ch.~5]{dolstraPurelyFunctionalSoftware2006} introduced
the input-addressed store for software deployment, where store paths encode
cryptographic hashes of all build inputs; ch.~6's intensional model extends
content-addressing to derivation outputs themselves. OxyMake adopts the
input-addressed principle for
workflow outputs: the cache key is a BLAKE3 hash of the rule source, input
content hashes, parameters, environment specification, and platform. The
key is input-addressed; the artefact store beneath it is content-addressed,
blobs being named by the BLAKE3 hash of their own bytes. One
distinction must be drawn: Dolstra's observation that ``if a build
succeeds, we know that we have specified all the dependencies''
(ch.~10) rests on the build taking place in a temporary directory using
only inputs already in the store, so that an incomplete declaration
fails loudly. OxyMake adopts the
content-addressed key but not that isolation: rules run as ordinary
processes, so the key is comprehensive over \emph{declared} inputs only,
and input completeness remains the user's obligation
(\S\ref{sec:undeclared-inputs}).

Neither axis of this taxonomy classifies what a rebuild decision
presupposes of its deployment, and among the systems compared here that
is the dimension on which OxyMake's default decision differs: Bazel and
Buck2 take a build daemon as their standard operating mode, Nix an
external store, GNU Make live timestamps in one working tree, and
Snakemake a provenance record bound to one working tree, while OxyMake, like DVC, presupposes nothing beyond the checkout
itself (\S\ref{sec:introduction}, \S\ref{sec:file-based}).

\subsection{Reproducibility and FAIR Workflows}
\label{sec:fair}

This subsection maps the
reproducibility property OxyMake delivers onto the existing FAIR workflow
literature, and marks the boundary between what the orchestration layer
records and what it delegates to the binary substrate.

\paragraph{A three-layer reading of Goble et al.}
Goble et al.~\cite{gobleFAIRComputationalWorkflows2020} argue that
computational workflows are first-class FAIR (Findable, Accessible,
Interoperable, Reusable) digital objects, not merely tools for producing
FAIR data: ``Workflows are research products in their own right,
encapsulating methodological know-how that is to be found and published,
accessed and cited, exchanged and combined with others, and reused as
well as adapted.'' They observe that a workflow exists in several
\emph{forms}---a specification with exemplar data, an implementation of
that design in a workflow management system, an instantiation of that
implementation ready to run with inputs and parameters bound, and a run
result with data products and provenance logs---and that each form may
carry different FAIR criteria. We adopt a three-layer reading of those
forms---\emph{abstract workflow} (the rule graph, independent of any
instance), \emph{concrete workflow} (the resolved DAG bound to specific
inputs), and \emph{execution trace} (the per-run provenance
record)---and treat each layer as requiring its own FAIR compliance.
The labels and the collapse to three layers are ours, not Goble et
al.'s. That reading is what OxyMake's three-graph architecture
(\S\ref{sec:three-graphs}) implements. The mapping is direct:
\code{RuleGraph} carries the abstract workflow,
\code{JobGraph} (post-resolution, post-cache pruning) carries the
concrete workflow, and \code{ExecGraph} together with the audit-state
SQLite tables and the NDJSON event stream carries the execution
trace.

\paragraph{The Wilkinson 2025 principles.}
Goble et al.\ conclude that FAIR principles for data and for software
``need to be extended in order to address the processual nature of
workflows'' and that ``new FAIR indicators will also need to be
developed''; they define none.
Wilkinson et al.~\cite{wilkinsonApplyingFAIRPrinciples2025} supply that
extension, restating the FAIR principles as a numbered set specific to
computational workflows (their Table~1). We refer to those numbered
entries as indicators throughout; Wilkinson et al.\ call them
principles.
Chue Hong et al.~\cite{chueHongFAIR4RSPrinciples2022}
complement this with the FAIR4RS principles for research software,
which apply to the OxyMake binary itself as a research artefact.
The two papers together define the
acceptance criteria a workflow engine must meet to be considered
FAIR-native; we report against them in
Table~\ref{tab:fair} and discuss the residuals in
\S\ref{sec:fair-eval}.

\paragraph{The four-layer FAIR reproducibility ladder.}
Inside the three-layer reading above there is a finer-grained ladder
that is useful for placing OxyMake in the existing landscape.

\begin{table}[htbp]
\centering\small\setlength{\tabcolsep}{4pt}
\begin{tabular}{@{}lll@{}}
\toprule
\textbf{Layer} & \textbf{Witness} & \textbf{Tool examples} \\
\midrule
L1 -- Substrate    & Same input hash $\Rightarrow$ same path$^{\dagger}$ & Guix store, Nix store,
                                                                 Docker digest \\
L2 -- Orchestration & Same plan hash $\Rightarrow$ same DAG       & OxyMake \code{ox.lock} \\
L3 -- Execution    & Same DAG $\Rightarrow$ same output hashes$^{*}$ & OxyMake content-cache,
                                                                 Bazel actions \\
L4 -- Audit        & Run trace decoupled from code               & OxyMake NDJSON +
                                                                 \code{state.db}, RO-Crate \\
\bottomrule
\end{tabular}
\caption{A four-layer FAIR reproducibility ladder. Each layer
  provides an independent witness, and the witnesses
  \emph{compose} rather than overlap. A workflow runner can provide the
  L2--L4 witnesses \emph{without} providing L1, provided it does not
  report the substrate's guarantees as its own.
  $^{*}$The L3 witness holds only for deterministic rules on a fixed
  substrate. $^{\dagger}$An input-addressed store fixes the output
  \emph{store path}; bit-identical output bytes additionally require
  the build itself to be
  deterministic~\cite{nixosReproducibleBuilds}.}
\label{tab:fair-ladder}
\end{table}

\paragraph{Delegating the substrate layer.}
OxyMake's cache key includes an \emph{environment specification} (e.g.,
\code{requirements.txt} content hash, Docker image reference, Guix
manifest hash), but
treats the substrate's contract as an out-of-model axiom
(\S\ref{sec:named-invariants}, substrate boundary). The engine does not
attempt to verify that a \code{uv.lock} or a Guix manifest is itself
bit-reproducible; it records the hash and trusts the substrate to
deliver. This delegation is deliberate: it lets a FAIR-archival reader
reproduce the OxyMake-level witnesses (L2--L4) on any substrate that
honours the recorded hashes. The guix-cwl reference
workflows~\cite{prinsGuixCWLPipelines2018}
\emph{illustrate} one such substrate-composition pattern---a CWL
workflow run under a Guix-managed environment, with Guix providing L1 and
the workflow runner providing L2--L4; PiGx~\cite{wurmusPiGxReproducibleGenomics2018}
is the same pattern with Snakemake, rather than CWL, as the
orchestration layer. Whether OxyMake's
orchestration-level contract composes cleanly with such a substrate is
a design conjecture and a future direction
(\S\ref{sec:future-work}), not a property attested here.

The reproducibility crisis in computational
science~\cite{pengReproducibleResearchComputational2011,
stoddenEnhancingReproducibilityComputational2016} underscores the need for
workflow systems that guarantee deterministic re-execution---and equally
underscores the need for explicit scope-marking: a runner that reports the
substrate's guarantees as its own delivers a false signal when the substrate changes.
OxyMake's content-addressable caching provides a stronger guarantee than
timestamp-based systems: the same declared specification---rule
source, declared input contents, parameters, environment
specification, and shell---$\Rightarrow$ same cache \emph{key}, under
the same key-format version and on any machine of the same
OS/architecture with the same layout inside the workflow root
(Eq.~\ref{eq:cache-key}). The
cache key is content-derived, whereas verification of an existing
artefact depends on the selected policy
(\S\ref{sec:content-addressable}). And the guarantee holds \emph{only
at L2--L4}. L1 is delegated, by design, to the substrate of the
reader's choice.

\subsection{Distributed Execution}
\label{sec:distributed-related}

DAG-based workflow scheduling on heterogeneous distributed systems is a
well-studied problem. The HEFT (Heterogeneous Earliest Finish Time)
algorithm~\cite{topcuogluPerformanceeffectiveAndLowcomplexity2002} and its
variants provide efficient heuristics for task placement. Adhikari et
al.~\cite{adhikariSurveySchedulingStrategies2019} survey cloud scheduling
strategies, identifying the trade-off between makespan minimization and
cost optimization.

Apache Airflow~\cite{airflow} popularised DAG-based workflow orchestration for
data engineering, modelling pipelines as Python-defined task graphs with
pluggable executors.  Argo Workflows~\cite{argoWorkflows} applies the same DAG
model natively on Kubernetes, encoding steps as container specifications in
YAML.  Both systems excel at scheduling heterogeneous tasks across distributed
infrastructure. Neither, however, \emph{automatically} derives a
reusable result key from the content of all declared file inputs:
Airflow's graph is built by executing Python, so its structure is
available only after the DAG file has been run, and Argo's
YAML---declarative in form---names container images and command lines
without hashing the data flowing between steps. Argo does expose a
memoization mechanism (\code{memoize:}, since v2.10) whose key is
user-supplied, so a user may construct a content-derived key by hand;
there the contrast is the absence of an automatic
declared-file-content key, not an impossibility. Airflow exposes no
task-result memoization primitive at all: a scheduled task instance is
executed.

Frameworks like Ray~\cite{moritzRayDistributedFramework2018} and
Dask~\cite{rocklinDaskParallelComputation2015} provide distributed
execution with task-level parallelism, but their principal surface is a
programming API rather than a declarative file-rule model: Dask is
Python-native, and Ray, whose 2018 paper introduces its task and actor
model, has since added Java and C++ APIs---both surfaces still expose
tasks and actors rather than file rules. OxyMake
bridges this gap
with its executor backends: the same declarative workflow file is
accepted on a local machine
(\code{-j~N}), a Slurm cluster (\code{--executor slurm}), or Ray
(\code{--executor ray}) without modification; a Kubernetes executor is
designed but not yet implemented (\S\ref{sec:feature-matrix}). Neither
distributed path is exercised in this evaluation, and both require the
workspace---state database included---on storage visible to the
submission and compute nodes, a configuration the local-disk
requirement of \S\ref{sec:distributed} does not support.

\section{Design Principles}
\label{sec:design}

OxyMake's central design contract is convergence:

\begin{quote}
\emph{\code{ox run} reconciles declared outputs against the current
tree. It runs the jobs that are missing or out of date and no others,
and is safe to repeat.}
\end{quote}

\noindent Two supporting principles make that contract auditable and
transportable, and both are written down with distinct scopes.
The first fixes what the engine attests and what it delegates: \emph{the
engine records the orchestration-level witnesses (L2--L4) and delegates
L1 to the substrate} (\S\ref{sec:fair}, Table~\ref{tab:fair-ladder}). The second fixes
the language boundary: \emph{Rust provides the engine, the workflow
provides the intent}. This is a separation of concerns: the engine
handles DAG resolution, scheduling, caching, and execution mechanics; the
workflow definition declares rules, dependencies, and resource requirements.
The engine applies no workload-dependent heuristics, and the
workflow specifies no scheduling logic and no cache management. The
engine does carry fixed operational constants---for example the
five-minute heartbeat age at which \code{ox clean} treats a session as
stale (\S\ref{sec:idempotent}). Each of
the principles below supports one or more cells of the reproducibility
ladder (Table~\ref{tab:fair-ladder}).

\paragraph{Three-layer mapping.}
Each of the six following principles is annotated with the layer(s) of
our three-layer reading of Goble et
al.~\cite{gobleFAIRComputationalWorkflows2020} (\S\ref{sec:fair}) that
it supports:
A = abstract workflow, C = concrete workflow, T = execution trace.
Content-addressable caching (\S\ref{sec:content-addressable})
supports C+T; daemon-free execution
(\S\ref{sec:idempotent}) supports T;
the machine-readable interface (\S\ref{sec:agent-api}) supports T's
machine-readability; the executor backends support A's portability;
named invariants and formal specifications
(\S\ref{sec:named-invariants}) support the trustworthiness of the
T-level audit trail; the declarative TOML
choice supports A's static parseability.

\subsection{Content-Addressable by Default}
\label{sec:content-addressable}

The cache \emph{key} is derived from file content, not timestamps;
whether an unchanged file is re-read or trusted on an unchanged
\code{(mtime,size)} pair is the validation policy described below. The
cache key for a job is:

\begin{equation}
\label{eq:cache-key}
k = \text{BLAKE3}\!\left(
  v \mathbin\| h_{\text{rule}} \mathbin\| h_{\text{inputs}}
  \mathbin\| h_{\text{params}} \mathbin\| h_{\text{env}}
  \mathbin\| h_{\text{shell}} \mathbin\| p
\right)
\end{equation}

\noindent where $v$ is a key-format version tag (bumping it cleanly
invalidates caches written under an older format), $h_{\text{rule}}$ is
the hash of the rule's source (command, inline code, script reference,
or function reference), $h_{\text{inputs}}$ is the sorted sequence of
$(\text{path}, \text{content hash})$ pairs covering declared inputs,
parameter files, and---in script mode---the script file itself,
$h_{\text{params}}$ captures parameter values, $h_{\text{env}}$ captures
the environment specification by \emph{content} (e.g., the bytes of the
referenced \code{requirements.txt} or conda YAML, not just its path),
$h_{\text{shell}}$ is the configured shell executable, and $p$ encodes
the platform (OS + architecture). Every component is length-framed with
a domain-separation tag, and optional components carry explicit presence
tags, so the pre-hash encoding is unambiguous: two distinct job
specifications serialize to two distinct byte streams. The key itself is
a 256-bit BLAKE3 digest of that stream, so distinct streams collide with
the cryptographic hash's negligible but non-zero probability. Binding
each content hash to its path prevents two inputs from exchanging
contents without changing the key. Because $p$ is part of the key, cache entries are shared only
between machines of the \emph{same} platform; heterogeneous OS/arch
cache reuse (e.g., develop on macOS arm64, run on Linux x86\_64) never
produces a false hit---and never hits at all---and is left as future
work. Each input's \emph{path} is hashed together with its content,
which binds a hash to the file it was computed for but also makes path
spelling part of the key. Paths are therefore recorded \emph{relative
to the workflow root}---defined as the directory \code{ox} is invoked
from, not the parent of the file passed via \code{-f}. Before entering
the key, a relative path is interpreted from that root, \code{.} and
\code{..} components are resolved lexically, and paths that exist on
disk are additionally canonicalised so that a symlink pointing outside
the root is seen for what it is. A path that lands \emph{inside} the
root after this normalisation is recorded relative; one that escapes
it---via an absolute spelling, a \code{..} prefix, or a symlink---is
recorded as a normalised absolute path, because its location is part of
the workflow's dependency and is machine-specific. Two checkouts of the
same tree at different absolute locations thus produce identical keys
for in-root inputs, which makes the key independent of the root's
absolute location and therefore identical across machines.
Key identity requires the same OS/architecture and the same layout
\emph{within} the workflow root, not the same location of that root. Two residual exclusions remain by design and are documented in
\S\ref{sec:limitations}: \code{call}-mode function bodies (the referenced
module is content-tracked only if declared as an input) and mutable
container image tags (hashed as written, not resolved to digests).

Cache validation is \textbf{pluggable}: users choose one of
three strategies through the \code{--cache-validation} option:

\begin{itemize}
\item \code{mtime+hash} (default) --- an unchanged \code{(mtime,size)}
  pair is trusted; if mtime or size differ, compute the
  BLAKE3 hash before declaring a hit or miss.  Fast on steady-state,
  correct on change: same-size corruption with a newer timestamp is
  detected rather than served.  The guarantee is bounded by what it
  watches---a rewrite that preserves both size and mtime never triggers
  the hash and is served as a hit, and this applies both to a
  previously stored input hash and to an existing output.  This policy
  addresses corruption that changes size or timestamp; \code{hash}
  below also covers rewrites that preserve both.
\item \code{mtime} (opt-in) --- pure filesystem metadata (stat calls only).
  Matches Make-style metadata-only behavior. The \emph{lookup} takes a
  stateless filesystem comparison with no dependency on
  \code{.oxymake/}; recording a completed job still writes the cache
  entry to the state database in every mode. No-op run time scales
  with workflow size: 313\,ms at 101 jobs and 500\,ms at
  10{,}001 jobs (the harness's \code{warm-mtime} rows at those two
  scales; the 10{,}001-job figure is quoted in \S\ref{sec:e2e-bench}). Content is never verified, so this
  mode is unsuitable for shared or multi-user caches.
\item \code{hash} --- always compute BLAKE3 hashes, unconditionally
  re-reading every byte.  The policy to use for CI reproducibility
  audits, and the one a shared or remote cache would require.
\end{itemize}

\noindent The strategy is configurable per invocation (\code{--cache-validation}),
per project (\code{[config] cache\_validation} in Oxymakefile.toml),
per environment (\code{OX\_CACHE\_VALIDATION}), or globally
(\code{\textasciitilde/.config/oxymake/config.toml}).
When a remote cache is in play (\code{ox run --cache-remote <dir>}, the
shipped directory backend), the validation strategy is promoted to
\code{hash} regardless of what was configured: a shared store has no
meaningful mtime relationship with the local workspace, so every
restored artefact is verified by content. The evaluated remote-cache
implementation is a shared-directory blob store: it stores
content-addressed artefact bytes, while the SQLite index that maps a
computation key to its output paths and hashes remains local to each
checkout, so restoring into a fresh checkout requires copying that
index separately. A computation-key-addressed manifest stored
remotely, which would make the directory backend self-sufficient, is
future work; S3 and GCS backends are not implemented
(\S\ref{sec:feature-matrix}).

Changing the serialized dimensions after release requires a bump of the
key-format version $v$, which invalidates every entry written under the
older format.

\paragraph{Threat model: undeclared inputs.}
\label{sec:undeclared-inputs}
Equation~\ref{eq:cache-key} is sound exactly over what it hashes:
\emph{declared} inputs. OxyMake executes rules as ordinary processes,
with no sandbox or syscall-level isolation. Anything a rule reads
without declaring it---a helper script invoked by the shell command, a
binary on \code{\$PATH}, a configuration file, a locale setting, an
environment variable outside the declared environment
specification---never enters the key, so a change to it produces a
\emph{false cache hit}: the engine reuses a stale output while
reporting that nothing changed. This is
the inverse of the phantom re-run, and it is the more dangerous
failure because it is silent. Nix and Guix close this hole with a
build sandbox that makes undeclared reads fail at build time; OxyMake
deliberately does not (rules must run unmodified user code on hosts
where namespace isolation is unavailable or unwanted), so the
completeness of the input declaration is trusted, not enforced. The
practical mitigations are discipline (declare scripts and tools as
inputs; pin the environment via \code{uv.lock} or an image digest,
which folds it into $h_{\text{env}}$) and audit (the \code{ox.lock}
plan of record makes the declared key inputs inspectable). Sandboxed
execution as an opt-in enforcement layer is future work
(\S\ref{sec:limitations}). Consequently, cache correctness assumes
complete input declarations.

\subsection{Daemon-Free Execution}
\label{sec:daemon-free}

No daemon process is required. Each \code{ox run} invocation is
self-contained: it resolves the DAG, executes jobs, writes state to
\code{.oxymake/state.db} (SQLite), and exits. The state layer
implements a persistent claim-state protocol of optimistic-lock SQL
transitions specified by \code{Cooperative\allowbreak Claim.tla}
in TLA+~\cite{lamportSpecifyingSystems2002}: its invariant
\textbf{INV-2} states that no two sessions hold the same job and that a
stale session's lease is reclaimed (\S\ref{sec:named-invariants}). This
protocol is implemented and model-checked but is not yet used as the
scheduling gate. In consequence, two \code{ox run} processes whose job
sets overlap currently execute the shared jobs independently rather than
one claiming and the other waiting; this is safe under the paper's
standing assumption that a declared rule is safe to re-run, and state
writes are atomic. No long-running control-plane service is required. An
optional \code{ox serve} mode is available for long-running
orchestration but is never required for a local run.

\subsection{Command and Library Interfaces}

Every operation is available as a CLI command with \code{--json} output,
so the same invocation serves a human and a program. The engine's
operations also live in library crates callable from Rust; a
single-facade crate (\code{ox-api}) exposes them through a typed entry
point (\code{Session\allowbreak Builder}, \code{plan}, \code{run}). The
facade is pre-1.0 and its surface may change; the command surface is
the interface this paper evaluates:

\begin{lstlisting}[style=shell]
# Human-readable
ox run results/all.vcf.gz -j 8

# Machine-readable (same operation)
ox run results/all.vcf.gz -j 8 --json
\end{lstlisting}

\subsection{Executor Backends}
\label{sec:distributed}

The same workflow file is accepted by every backend; moving from a
laptop to a cluster changes one flag, not the workflow. Three executor backends are
implemented---local, Slurm, and Ray---where the local backend runs
sequentially by default and in parallel with \code{-j N}; the Kubernetes
executor is designed but not yet
implemented (\S\ref{sec:feature-matrix}):

\begin{table}[htbp]
\centering\small
\begin{tabular}{@{}lll@{}}
\toprule
\textbf{Level} & \textbf{Flag} & \textbf{Mechanism} \\
\midrule
Sequential  & (default)           & Single process \\
Local par.  & \code{-j N}       & Tokio thread pool \\
Slurm       & \code{--executor slurm} & \code{sbatch}/\code{sacct} \\
Ray         & \code{--executor ray}   & Ray Jobs API \\
Kubernetes  & \code{--executor k8s}   & K8s Job via kube-rs \emph{(not yet implemented)} \\
\bottomrule
\end{tabular}
\end{table}

\noindent All entries require the \code{.oxymake/} state directory on
local disk. It is created in the process working directory and cannot be
relocated, so in practice the requirement is on the \emph{workspace}: the
directory \code{ox run} is invoked from must be on local disk, together
with the inputs and outputs resolved relative to it. On a cluster the
scheduler runs on the submission node and holds the SQLite state there;
compute nodes run jobs and never touch the database. Multi-node
coordination across NFS, Lustre, or GPFS, and coordination across
multiple submission nodes, are future work (\S\ref{sec:limitations}).
Because the Slurm and Ray backends need the workspace---inputs and
outputs---visible to the compute nodes, while \code{.oxymake/} cannot
be placed apart from that workspace, neither has a
supported deployment configuration in v0.1.0. The evaluation is
single-machine; no Slurm or Ray run is measured
(\S\ref{sec:evaluation}).

\subsection{Named Invariants and Formal Specifications}
\label{sec:named-invariants}

We use bounded model checking to examine safety properties of
concurrent state transitions; this discipline backs the L4 audit-trail
witness of the FAIR ladder (Table~\ref{tab:fair-ladder}).

OxyMake's runtime is built from components that are individually
deterministic---a worker, a scheduler, an evictor, a cancel path, a
state database---yet whose composition is concurrent. A FAIR-archival
reader has no way to distinguish ``the workflow reproduced'' from
``the workflow ran once on a single-session, no-fault path that
happened to produce the same bytes.'' The discipline that lets us
\emph{claim} reproducibility on the multi-session, fault-tolerant
path is formal specification of the cross-session safety properties.
Without it, the L4 audit trail is a log of one history out of an
unknown number of possible histories.

Concurrent sessions may fail independently while updating shared
state; we therefore model claim, cancellation, and recovery
transitions jointly. On at least three surfaces (multi-session
reclaim, cancel propagation, and post-crash recovery) the state of the
system is a \emph{relation} between the states of peers that can each
fail independently. The reachable interleavings on those surfaces
far outnumber what any feasible test suite can sample; a green CI
corroborates a very small fraction of the state space.

The structural reference is Newcombe et
al.~\cite[pp.~66--73]{newcombeHowAmazonWebServices2015}, which reports
applying TLA+ to ten large production systems at Amazon and tabulates
the bugs model checking found in six components of four of them (S3,
DynamoDB, EBS and an internal distributed lock manager)---bugs that,
in the cases the article describes, had passed design review, code
review and testing. Those systems were orchestrated on deterministic
components; the hazard class was concurrent, not adversarial. The
structural similarity to \code{ox run} is what justifies the
technique here, not any claim about input non-determinism.

\paragraph{Scope and ratio.}
OxyMake ships three TLA+ specifications totalling 623~lines (\code{.tla}
source: \code{Cache\allowbreak Consistency.tla} 201~L, \code{Cooperative\allowbreak Claim.tla}
254~L, \code{Cancel\allowbreak Propagation.tla} 168~L). What these specifications
verify, and at what bounds, is stated in Table~\ref{tab:verified-vs-assumed}
and Appendix~\ref{app:reproducibility}; that scope, not any line count,
is the measure of the effort. Against \metricSLOC~SLOC of Rust workspace
code the specifications are 0.96\% of the source; the figure is
descriptive context, not a claim of coverage. Post-crash
recovery is outside the evaluated scope.

\paragraph{Mapping invariants to specifications.}
Each specification checks one or more named invariants, derived from the
properties the implementation depends on:

\begin{table}[htbp]
\centering\small
\begin{tabular}{@{}lll@{}}
\toprule
\textbf{Invariant} & \textbf{Property} & \textbf{Spec} \\
\midrule
\textbf{OX-1} & cache key determinism                          & \code{Cache\allowbreak Consistency.tla} \\
\textbf{OX-2} & no backchannel (info direction)                & --- (architectural, not model-checked) \\
\textbf{OX-6} & stationary cache safety                        & \code{Cache\allowbreak Consistency.tla} \\
INV-3a        & \code{CancelledNeverCached}                     & \code{Cancel\allowbreak Propagation.tla} \\
INV-2         & claim atomicity / stale-session reclaim         & \code{Cooperative\allowbreak Claim.tla} \\
INV-3b        & \code{JobFailedImpliesNoIntent}                 & \code{Cancel\allowbreak Propagation.tla} \\
INV-3c        & \code{No\allowbreak Zombie\allowbreak Running}                  & \code{Cancel\allowbreak Propagation.tla} \\
\bottomrule
\end{tabular}
\caption{Named invariants and the specifications that check them.
  OX-1, OX-2, and OX-6 name product-level properties; INV-2 and
  INV-3 name the concurrent safety properties checked by TLC.}
\label{tab:named-invariants}
\end{table}

\paragraph{Bounds.}
``Model-checked'' here is a claim about \emph{bounded
model checking}, not proof. TLC exhaustively explores the reachable
state space of each spec at small, fixed instance sizes:
\code{Cache\allowbreak Consistency} at 2 workers $\times$ 3 rules,
\code{Cooperative\allowbreak Claim} at 3 sessions $\times$ 2 jobs (TTL 2, clock
bound 4), \code{Cancel\allowbreak Propagation} at 3 jobs. Within those bounds
the invariants hold over \emph{every} interleaving---this is the
qualitative step beyond testing, which samples interleavings. Beyond
them the invariants are corroborated, not verified; no inductive proof
(TLAPS) has been attempted.

\paragraph{Assumptions.}
The specs import axioms they do not check (next paragraph); the
model-checked guarantee is conditional on those axioms holding in
deployment. In addition, OX-1 as model-checked is conditional on key
purity, and that condition is itself modelled: in
\code{Cache\allowbreak Consistency.tla} each materialisation draws the
rule's key from a constant set \code{Key\allowbreak Variants}, and a
worker crash before registration lets a peer recompute the key for the
same rule. Under the shipped configuration
(\code{Key\allowbreak Variants = \{1\}} --- the key is a pure function
of the rule's declared inputs) \code{CacheKeyDeterminism} holds over
all crash/re-claim interleavings; a second configuration
(\code{Key\allowbreak Variants = \{1,2\}}, modelling an undeclared
input leaking into the key) refutes it, which is the evidence the
invariant has content rather than being true by construction. Purity
of the real BLAKE3 key remains an assumed property, and the
input-declaration completeness assumption
(\S\ref{sec:undeclared-inputs}) remains out of model.

\paragraph{Substrate boundary.}
The specifications import named axioms about SQLite, the filesystem,
the kernel, and the executor process; the right column of
Table~\ref{tab:verified-vs-assumed} enumerates them. These
assumptions are explicit and are not checked by the model.
One imported axiom deserves emphasis because it can fail in exactly
the deployment where multi-session execution is most likely to be
attempted---several sessions pointed at one workspace on a shared
filesystem:
\code{Cooperative\allowbreak Claim.tla} assumes
\textbf{\code{StateDbAtomicCommit}}---that a SQLite \code{COMMIT} is
all-or-nothing with respect to concurrent writers, so the
\code{reclaim\_stale\_jobs} transaction is atomic. SQLite delivers
that on a local filesystem with working POSIX locks; on NFS, Lustre,
or GPFS---shared filesystems where one might naturally point several
sessions at one workspace---file locking is unreliable and the premise
is \emph{false}. This is why OxyMake requires \code{.oxymake/} on
local disk (\S\ref{sec:limitations}): the requirement is what
discharges the axiom. It is a requirement on the operator, not a
mechanism the engine provides---\code{.oxymake/} is created in the
working directory and there is no way to place it elsewhere, so
satisfying it means putting the whole workspace on local disk. If the
state database is run on NFS, the premise fails and INV-2's
model-checked guarantee no longer applies.

\paragraph{Verified vs.\ assumed.}
Table~\ref{tab:verified-vs-assumed} draws the line in one place.

\begin{table}[htbp]
\centering\small
\begin{tabular}{@{}p{0.46\linewidth}p{0.46\linewidth}@{}}
\toprule
\textbf{Verified (TLC, bounded, exhaustive)} & \textbf{Assumed (imported, not checked)} \\
\midrule
\code{Done\allowbreak By\allowbreak Claim\allowbreak Holder},
  \code{At\allowbreak Most\allowbreak One\allowbreak Claimer}
  (INV-2; 3 sessions, 2 jobs; the zombie-terminal-write red
  configuration refutes the former when the session filter is
  disabled) &
\code{State\allowbreak Db\allowbreak Atomic\allowbreak Commit} --- SQLite commit atomicity;
  holds on local disk, \emph{false on NFS/Lustre/GPFS}; discharged by the
  local-disk requirement (\S\ref{sec:limitations}) \\
\addlinespace
\code{Cache\allowbreak Key\allowbreak Determinism},
  \code{Register\allowbreak Precedes\allowbreak Read},
  \code{No\allowbreak Double\allowbreak Register}
  (OX-1/OX-6; 2 workers, 3 rules, worker crashes; the
  nondeterministic-key red configuration refutes the first) &
\code{Executor\allowbreak Honest}, \code{User\allowbreak Code\allowbreak Matches\allowbreak Rule},
  \code{Storage\allowbreak Delete\allowbreak Atomic},
  \code{Executor\allowbreak Failure\allowbreak Classification}
  (substrate axioms) \\
\addlinespace
\code{Cancelled\allowbreak Never\allowbreak Cached},
  \code{Job\allowbreak Failed\allowbreak Implies\allowbreak No\allowbreak Intent},
  \code{No\allowbreak Zombie\allowbreak Running} (INV-3; 3 jobs) &
BLAKE3 key purity (a pure function of declared inputs ---
  \code{Key\allowbreak Variants = \{1\}} in the model) and input-%
declaration completeness (\S\ref{sec:undeclared-inputs}) \\
\bottomrule
\end{tabular}
\caption{What the TLA+ suite verifies versus what it assumes. The left
  column is exhaustively checked by TLC at the stated bounds; the right
  column is the conditional premise of every left-column guarantee.}
\label{tab:verified-vs-assumed}
\end{table}

The full specification inventory, with the bounds explored and the
state counts, is given in Appendix~\ref{app:reproducibility}.

\subsection{Declarative Workflow, Polyglot Execution}

The workflow definition is always TOML---declarative, statically parseable,
not Turing-complete. This is a deliberate departure from Snakemake's Python
DSL. However, individual rules execute in any language through four
execution modes forming a spectrum from opaque to optimizable
(Section~\ref{sec:exec-spectrum}).

\subsection{Errors as First-Class Design}

Every error message traces the full causal chain through the DAG: which job
failed, what the root cause was (exit code, OOM, missing input), which
upstream rule was responsible, and what corrective action is available. In
JSON mode, the same structure is machine-parseable, enabling downstream tools to
programmatically read error chains and take corrective action.

\section{Architecture}
\label{sec:architecture}

\subsection{The Three Graphs}
\label{sec:three-graphs}

OxyMake uses three distinct graph representations at different abstraction
levels, following the pattern proven by DataFusion (logical $\to$ physical
plan), Bazel (target $\to$ action graph), and Spark (RDD lineage $\to$
stages $\to$ tasks).

\paragraph{RuleGraph (logical).}
The workflow as declared by the user. Nodes are \code{Rule} objects with
unresolved wildcards; edges represent input/output pattern dependencies
between rules. The RuleGraph is compact and abstract: one \code{call}
node represents \emph{all} variant-call instances. Operations at this level
include cycle detection, rule ambiguity detection, and structural
validation. The hierarchical visualization command (\code{ox dag
--group-by stage}) operates on this representation.

\paragraph{JobGraph (physical).}
The RuleGraph expanded (wildcards resolved, conditional guards evaluated)
and then optimized through a series of passes. Nodes are
\code{ConcreteJob} objects with fully resolved wildcards; edges are
typed as \code{Produces}, \code{Consumes}, or \code{Blocks}. The
JobGraph wraps a \code{petgraph::DiGraph}~\cite{petgraph} with typed
node variants (\code{Job}, \code{Output}, \code{Gate}).

One optimization pass on the JobGraph is implemented: \emph{cache
pruning}, which marks jobs with up-to-date outputs as \code{Skipped}
using the content-addressable cache (Equation~\ref{eq:cache-key}). It is
the pass the evaluation exercises. Five further passes---task fusion,
materialization elimination, group scheduling, critical path analysis
and partition planning---are designed against the same trait interface
but not implemented (\S\ref{sec:limitations}).

\paragraph{ExecGraph (runtime).}
The JobGraph annotated with live execution state. Each node carries a status
(Pending $\to$ Ready $\to$ Running $\to$ Completed/Failed/Skipped),
runtime metrics (wall time, peak memory), and log paths. The scheduler
dispatches \code{Ready} nodes, the reporter emits events, and crash
recovery serializes the ExecGraph to SQLite for resumption.

\begin{figure}[htbp]
  \centering
  \includegraphics[width=0.75\textwidth,height=0.85\textheight,keepaspectratio]{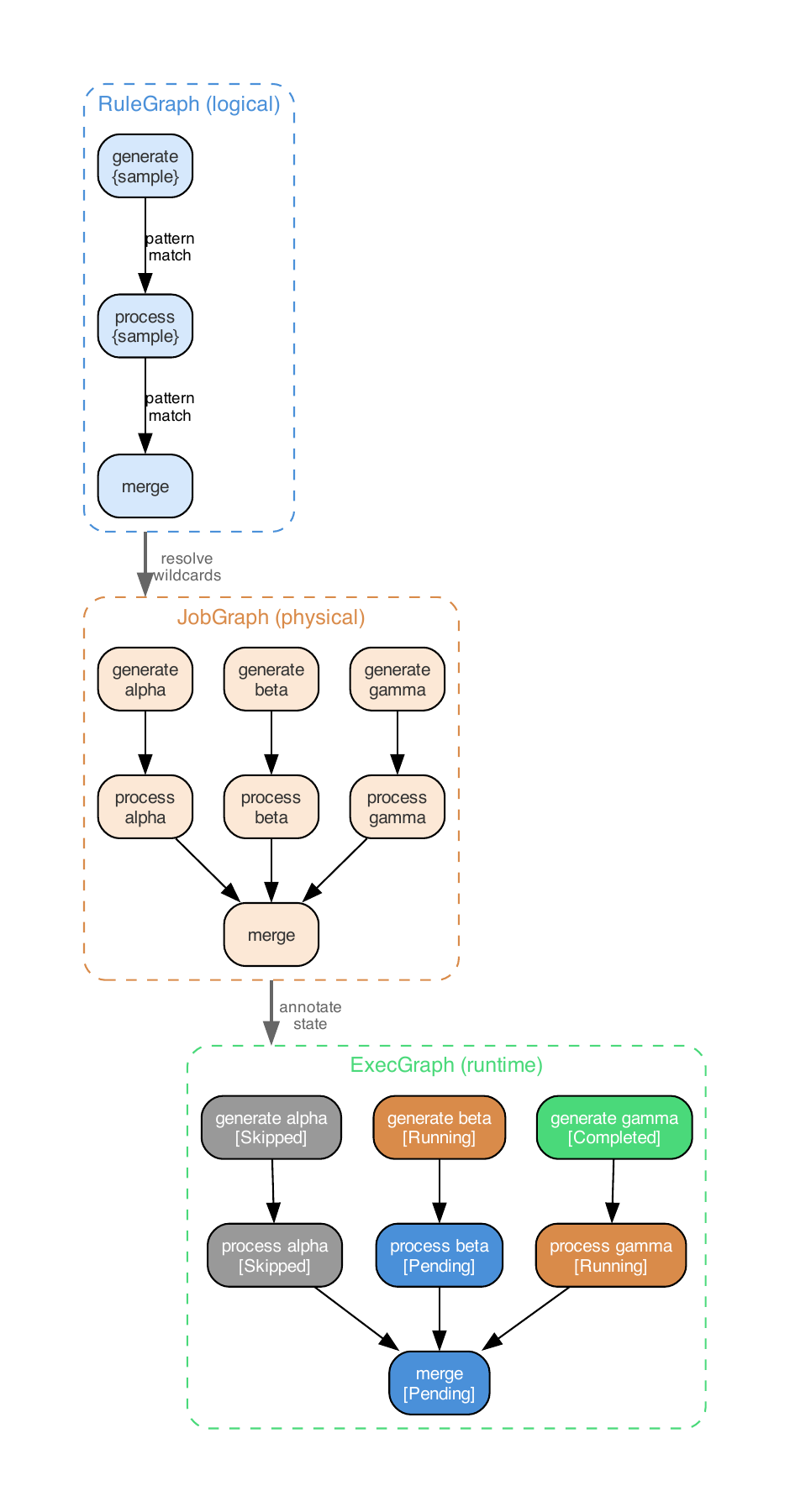}
  \caption{The three-graph architecture applied to the demo pipeline.
    \emph{RuleGraph} (top): abstract rules with unresolved wildcards.
    \emph{JobGraph} (middle): wildcards resolved to concrete instances
    (alpha, beta, gamma), revealing the true parallelism structure.
    \emph{ExecGraph} (bottom): annotated with runtime state---gray nodes
    are cache-skipped, orange are running, blue are pending, green are
    completed.}
  \label{fig:three-graphs}
\end{figure}

\begin{figure}[htbp]
  \centering
  \includegraphics[width=0.85\textwidth]{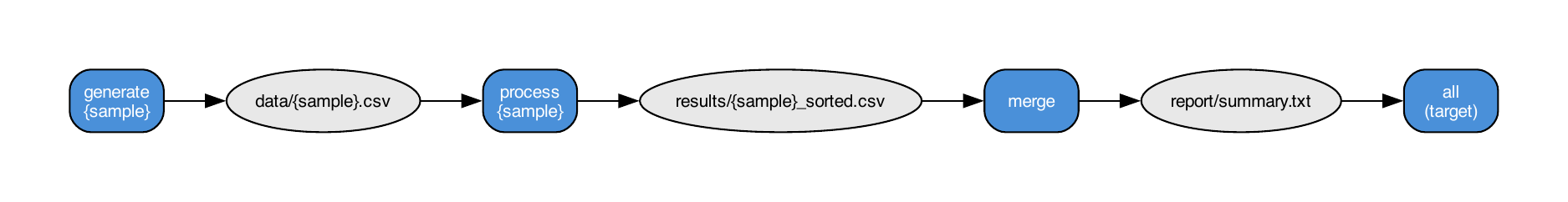}
  \caption{DAG visualization of the demo word-frequency pipeline generated
    by \code{ox dag --format dot}. Rule nodes (blue boxes) are connected
    through file pattern intermediaries (gray ellipses), forming a bipartite
    graph that makes data flow explicit. The pipeline progresses left to right:
    generate $\to$ process $\to$ merge $\to$ target.}
  \label{fig:demo-dag}
\end{figure}

The full pipeline is: \code{Oxymakefile.toml} $\to$ parse $\to$
\code{RuleGraph} $\to$ resolve wildcards $\to$ \code{JobGraph (raw)}
$\to$ optimization passes $\to$ \code{JobGraph (optimized)} $\to$
annotate with runtime state $\to$ \code{ExecGraph} $\to$ schedule and
execute. Figure~\ref{fig:three-graphs} illustrates this progression
using the demo word-frequency pipeline, and Figure~\ref{fig:demo-dag}
shows the RuleGraph as generated by \code{ox dag --format dot}.

\subsection{The Execution Spectrum}
\label{sec:exec-spectrum}

OxyMake provides four execution modes forming a spectrum from maximum
flexibility to maximum optimizability:

\begin{table}[htbp]
\centering\small
\begin{tabular}{@{}llll@{}}
\toprule
\textbf{Mode} & \textbf{I/O} & \textbf{Memory} & \textbf{Optimize} \\
\midrule
\code{shell}  & Files only & No  & Opaque \\
\code{run}    & Files only & No  & Limited \\
\code{script} & Files only & No  & Limited \\
\code{call}   & Files or mem & Yes & Full \\
\bottomrule
\end{tabular}
\caption{The four execution modes. The \textbf{Optimize} column rates
  the headroom a mode affords, not work the engine performs today: of
  the passes that would exploit it, only cache pruning is implemented
  (\S\ref{sec:three-graphs}).}
\label{tab:exec-spectrum}
\end{table}

The \code{call} mode is the most optimizable of the four. A rule
declares a function reference (e.g.,
\code{pipeline.features:compute\_features}) that the user is expected to
keep free of side effects; OxyMake manages all I/O outside the function.
The engine does not enforce this: the cache key covers the function
reference, not the function body, so editing the imported module without
declaring it as an input can serve a stale result
(\S\ref{sec:limitations}). Shell, \code{run}, and \code{script} modes
hash the rule body itself---the inline command or code, or the named
script file---so an edit to the body does change the key. Code that
the body \emph{invokes} is not covered in any of the four modes: a
helper script called from a shell command is an undeclared input like
any other (\S\ref{sec:undeclared-inputs}). The four modes differ in
what counts as the body, not in whether the undeclared-input hazard
applies.

\begin{lstlisting}[style=toml]
[rule.compute_features]
input = [{ path = "data/{sample}.parquet",
           format = "parquet" }]
output = [{ path = "features/{sample}.parquet",
            format = "parquet",
            materialize = "auto" }]
call = "pipeline.features:compute_features"
\end{lstlisting}

The Python function receives a DataFrame, returns a
DataFrame, and performs no file I/O:

\begin{lstlisting}[style=python]
def compute_features(df: pl.DataFrame) -> pl.DataFrame:
    return df.with_columns(
        anomaly=pl.col("value").rolling_mean(20),
        spread=pl.col("value").rolling_std(60),
    )
\end{lstlisting}

In file mode, OxyMake reads the input using the declared format codec,
calls the function, and writes the result. In memory mode, the transfer
mechanism depends on the executor: the Ray executor passes data through
the Ray object store (\code{ray.put}/\code{ray.get}), avoiding disk
entirely, while the local executor currently serializes intermediates
to disk via format codecs (Arrow IPC transport is planned).
The transfer mechanism is invisible to the function.

\paragraph{Materialization policy.}
The \code{materialize} field controls when outputs touch disk.
Four modes are available: \code{always} (default, reproducible),
\code{auto} (only if a non-\code{call} downstream needs the file),
\code{never} (memory only, not cached), and \code{final}
(only DAG leaves). A global override
(\code{ox\,run\,-{}-materialize=final}) switches the entire workflow
between modes, enabling rapid exploration with reduced disk overhead
(zero on executors with native object stores).

\paragraph{Language interop.}
OxyMake communicates with language runtimes via subprocess and
structured serialization rather than embedding (e.g., PyO3). This
design choice preserves compatibility with
isolated environments (\code{uv}, \code{conda}, Docker) and avoids
coupling to specific language ABIs. Worker processes are reused
across sequential \code{call}-mode jobs in the same environment to
amortize startup cost.

\subsection{Plugin Architecture}
\label{sec:plugins}

OxyMake defines five plugin axes, each specified by a Rust trait. Each
axis lists the backends implemented today; others named the trait admits
but which are not implemented are marked \emph{(planned)}:

\begin{description}
\item[Executor] Where jobs run: local, Slurm, Ray; Kubernetes \emph{(planned)}.
\item[Storage] Where files live: local filesystem; S3, GCS \emph{(planned)}.
\item[Environment] How jobs are isolated: system, uv, conda, Docker; Nix \emph{(planned)}.
\item[Reporter] How progress is communicated: terminal, NDJSON; webhooks \emph{(planned)}.
\item[FormatCodec] How objects are serialized: Parquet, CSV, JSON, and columnar formats.
\end{description}

\noindent One placement note reconciles this list with the crate map
(\S\ref{sec:crates}): the shipped environment support (system
passthrough, and command wrapping for \code{uv}, conda, and Docker) is
implemented inside the local executor crate
(\code{ox-exec-local}), and local file handling lives in the engine
crates. The dedicated \code{ox-env-*} and \code{ox-storage-local}
crates are extraction targets that today contain trait scaffolding
only.

Plugin selection is compile-time via Cargo feature flags. The minimal binary
(local executor + local storage + system environment + terminal reporter +
basic codecs) has minimal dependencies and compiles fast. The feature-flag
mechanism is where heavy plugins (Kubernetes, S3) will be gated when
implemented; today it keeps the default binary small by excluding
optional codecs and reporters.

The \code{Executor} trait is the primary extension point:

\begin{lstlisting}[style=rust]
trait Executor: Send + Sync {
    async fn execute(
        &self, job: &ConcreteJob,
        ws: &Workspace, ctx: &ExecContext
    ) -> Result<JobResult>;
    async fn cancel(&self, id: &JobId)
        -> Result<()>;
    fn capabilities(&self)
        -> ExecutorCapabilities;
}
\end{lstlisting}

The \code{capabilities()} method reports whether the executor supports GPU
scheduling, streaming between co-scheduled jobs, hermetic workspace
isolation, and in-memory passing between \code{call}-mode jobs. When
an executor does not support memory passing (e.g., the local executor's
current disk-serialized path, Slurm, Kubernetes), the scheduler
automatically promotes \code{InMemory} outputs to \code{File}. A
workflow whose in-memory values are representable in the configured
serialization format therefore runs on backends without a native
object store, at the cost of disk round-trips; values the format
cannot represent, and the semantics of
\code{materialize\,=\,"never"} under promotion, are limitations stated
in \S\ref{sec:limitations}.

\subsection{Idempotent Convergent Execution}
\label{sec:idempotent}

\code{ox run} ensures that the declared outputs exist.
This is a declarative, convergent model inspired by
\code{terraform apply}, formalized as a state-machine reconciliation in
the sense of Lamport~\cite{lamportSpecifyingSystems2002}. Every invocation
reconciles desired state versus actual state:

\begin{table}[htbp]
\centering\small
\begin{tabular}{@{}lp{0.58\linewidth}@{}}
\toprule
\textbf{Current State} & \textbf{Action} \\
\midrule
Output cached, inputs unchanged & Skip \\
Output missing (intermediate deleted) & Re-execute + cascade \\
Job running (another session) & Re-execute (duplicated work; no claim gate today,
  and concurrent writers to one output path are not serialised, \S\ref{sec:daemon-free}) \\
Job pending & Execute \\
Job failed previously & Re-execute \\
\bottomrule
\end{tabular}
\end{table}

\paragraph{Transitive invalidation of deleted intermediates.}
The ``output missing'' row states a policy rather than a derived
property: when a recorded output is missing, OxyMake re-executes its
producer and, conservatively, that producer's transitive dependents,
rather than re-establishing the dependents' keys from the regenerated
bytes. When a user deletes an intermediate output
file (e.g., \code{rm results/merged.parquet}), OxyMake's cache pre-scan
detects the absence: the cache store verifies that every recorded output
\emph{still exists on disk} before declaring a cache hit. A missing file
invalidates the producing job's cache entry and triggers \textbf{transitive
cascade}---all downstream jobs whose inputs transitively depend on the
deleted file are marked stale, regardless of whether their own direct
inputs appear intact.

This cascade is implemented by propagating staleness through topological
traversal of the job graph: if any upstream job must re-execute, all its
direct dependents are marked stale, and this propagation continues until
all transitively affected jobs are identified. The propagation is
unconditional---it does not re-check whether a regenerated input's hash
still matches---so on this path the engine behaves as a topologically
propagated dirty bit rather than as the verifying-traces rebuilder of
\S\ref{sec:build-theory}.

Snakemake~7.32.4 evaluates each rule against its \emph{direct}
inputs and outputs. In some configurations, if an intermediate
file is deleted but a downstream output still exists,
Snakemake may report ``Nothing to be done.'' Snakemake~7+'s recorded
provenance (\texttt{-{}-rerun-triggers}) governs re-execution on code,
parameter, and input-set changes, but the existence check remains
per-rule and local.
OxyMake's cache pre-scan avoids this issue: it verifies that
every declared output still exists on disk, so a deleted intermediate
always triggers a transitive rebuild. For example, after deleting a
single intermediate file in a multi-stage pipeline, \code{ox run} rebuilds
that stage and all downstream stages (preprocess $\to$ merge
$\to$ analyze $\to$ report). A minimal reproduction comparing both
tools is in the public repository under
\code{examples/\allowbreak intermediate-deletion/}.
The property this cascade enforces---no downstream job observes an
output whose producer has been invalidated but whose cache entry has
not been retired---is defended by the implementation's cache pre-scan
and its integration tests; a dedicated specification of the eviction
race is outside the evaluated scope
(\S\ref{sec:named-invariants}).

\noindent Concurrent sessions on \emph{disjoint} job sets are the
supported pattern today:

\begin{lstlisting}[style=shell]
# Terminal 1: dataset-A pipeline
ox run --where dataset=train

# Terminal 2: dataset-B pipeline (concurrent)
ox run --where dataset=test
\end{lstlisting}

For \emph{overlapping} job sets, the state layer implements a
persistent claim-state protocol: atomic SQL updates with optimistic locking
(an \code{UPDATE\ldots WHERE status='pending'} that affects zero rows
means another session claimed the job first), session heartbeats, and a
two-phase reclaim that resets the jobs of stale sessions to pending.
The staleness threshold is a parameter of the reclaim call, not an
engine-wide constant; the shipped engine invokes the reclaim only from
\code{ox clean}, at a heartbeat age of five minutes.
\code{Cooperative\allowbreak Claim.tla}
model-checks this protocol, which is not yet the scheduling gate
(\S\ref{sec:daemon-free}).

Five commands share the convergent model and one filter surface:
\code{ox run} (converge), \code{ox cancel} (abort), \code{ox
invalidate} (reset), \code{ox plan} (preview), and \code{ox status}
(observe). All five commands accept \code{--where}, \code{--rule}, and
\code{--json} flags.

\subsection{Tags, Guards, and Incremental Growth}
\label{sec:tags}

Large, evolving workflows are managed through tags, conditional
guards, and incremental growth.

\paragraph{Tags.}
Every resolved wildcard automatically becomes a tag (implicit tags).
Rules can also declare explicit tags for cross-cutting concerns
(e.g., \code{stage\,=\,"features"}, \code{cost\,=\,"high"}).
The \code{-{}-where} filter selects target jobs by tag values,
then backward-chains to include all necessary upstream dependencies.
Tags also enable \textbf{hierarchical DAG visualization} via
\code{ox\,dag\,-{}-group-by\,stage}, which collapses jobs into
meta-nodes by tag grouping---keeping very large DAGs comprehensible
at a glance.

\paragraph{Conditional guards.}
The \code{when} clause enables non-uniform DAG branches where some
wildcard instances need sub-workflows that others do not:

\begin{lstlisting}[style=toml]
[config]
high_variance_samples = ["s02", "s07", "s11"]

[rule.spectral_analysis]
input = ["results/{sample}.parquet"]
output = ["diagnostics/{sample}/outliers.png"]
when = "sample in @high_variance_samples"
\end{lstlisting}

Guards are evaluated at DAG resolution time---a job whose guard is false
is never created in the graph. This enables instance-specific
branching without phantom nodes. The domain is correspondingly
restricted: a guard expression ranges over resolved wildcard bindings,
\code{[config]} scalars and lists (the \code{@name} form above),
environment variables, and file existence---all readable before
resolution. It cannot name a value the workflow itself computes, so a
set such as \code{high\_variance\_samples} must be declared in
\code{[config]}, as above, or written there by a prior run.
Because the guard is evaluated before job construction, guard state
does not enter Eq.~\ref{eq:cache-key}: a job that is never created has
no key. The environment-variable and file-existence forms therefore
make resolution depend on live, unhashed state.

\paragraph{Incremental growth.}
Research workflows grow iteratively: explore, select, evaluate, refine.
OxyMake supports this through content-addressable incrementality (a new
rule does not change the key of any job whose own declared
specification is unchanged; it can still change which rule produces a
path, or make resolution ambiguous), workflow composition via
\code{include} directives, snapshots for milestone comparison
(\code{ox snapshot diff baseline-v1}), and run annotations
(\code{ox run --note "Testing new preprocessing step"}) recorded in the
state database alongside each run's metrics.

\section{Implementation}
\label{sec:implementation}

\subsection{Rust as Implementation Language}
\label{sec:rust}

Rust was chosen because its properties directly serve OxyMake's design goals.

\paragraph{Type system as architecture.}
Rust's type system enforces architectural invariants at compile time.
\code{Rule} (unresolved wildcards) and \code{ConcreteJob} (fully resolved)
are distinct types, so an unresolved rule cannot be passed where a
scheduled job is expected. \code{Send + Sync} bounds rule out data
races on shared state at compile time; the protocol-level concurrency
hazards they do not cover---deadlock, lost updates, and torn multi-step
protocols across sessions---are the subject of
\S\ref{sec:named-invariants}. \code{Result<T, E>}
on fallible functions makes ordinary error propagation explicit rather
than exceptional, so a failing job returns an error the scheduler
handles instead of unwinding the process. The scheduler cancels
in-flight jobs on shutdown or panic, sending \code{SIGTERM} to each
process group to clean up child and grandchild processes; a
\code{SIGKILL} to the scheduler itself, or a kernel-level failure,
bypasses this path.

\paragraph{Performance ceiling.}
Three properties bound the engine's speed. The workflow format is
inert: it is parsed as data rather than executed, where a Python-DSL
workflow incurs interpreter import time on every invocation. The graph algorithms
are asymptotically cheap: the \code{petgraph} library~\cite{petgraph}
provides an $O(|V|+|E|)$ depth-first topological sort
(\code{algo::toposort}, the API the engine calls), and the
\code{ProducerIndex} pays a one-time $O(P)$ pattern compilation in the
output-pattern count $P$, after which each target lookup is a linear
scan over the $P$ precompiled patterns---$O(T \times P)$ to resolve $T$
targets (\S\ref{sec:scale-study}).
The measured range extends to 10{,}001 jobs
(Table~\ref{tab:dag-resolution}); extrapolating the observed near-linear
slope to $5\times10^4$ and $10^5$ jobs is a projection, not a
measurement (\S\ref{sec:scale-study}). We expect hashing not to
be the bottleneck: BLAKE3~\cite{blake3,blake3spec} is reported at
6{,}866~MiB/s ($\approx$6.7~GiB/s) single-threaded on x86-64
(Cascade Lake-SP, 16~KiB input),
above the I/O rate of the files a typical workflow reads. The evaluated
host is arm64, where BLAKE3 takes a different SIMD path and the figure
is untested. We have not
profiled the hash phase of our benchmark, so this expectation rests on
published throughput figures for another architecture, not on a
measurement of this system.

\paragraph{Distribution.}
OxyMake ships as one self-contained \code{ox} executable: it bundles no
interpreter and requires no OxyMake daemon or service, linking only the
host platform's system libraries (we do not claim a fully static link,
which would require a per-target linkage attestation we have not
published). Cross-compilation targets Linux, macOS, and Windows. Installation requires
only \code{cargo install oxymake} or downloading a pre-built binary.
The workspace compiles via standard \code{cargo build --release}.

\subsection{Testing and Documentation}
\label{sec:methodology}

\paragraph{Testing.}
\metricTests~tests accompany \metricSLOC~SLOC
(Table~\ref{tab:crate-size}); test density varies widely between the
engine crates and peripheral ones. The
testing stack is \code{cargo test} for unit and doc tests,
\code{cargo-llvm-cov} for line coverage, \code{proptest} for
property-based testing of DAG invariants and cache-key stability, and
\code{insta} for snapshot testing of TOML parsing and CLI output.

\paragraph{Literate programming.}
Public functions and types carry \code{///} doc comments with executable
examples that compile and run as part of \code{cargo test}. The
\metricDocTests~doc tests are documentation, usage examples, and
regression tests at once. Because they run in CI, a documented example
that stops compiling fails the build, which bounds---rather than
eliminates---documentation drift for the examples under test. Prose
documentation and design rationale in module headers and Architecture
Decision Records (ADRs) are not machine-checked. The project includes
\metricDocFiles~documentation files covering concepts, command
references, format specifications, and error indices.

\subsection{Crate Architecture and the Cache Control Path}
\label{sec:crates}

OxyMake is organized as a Cargo workspace of \metricCrates~crates (24
workspace members plus \code{ox-metrics}, which is workspace-excluded
for dependency isolation), each with a single, well-defined
responsibility. The per-crate census and the dependency graph are in
Appendix~\ref{app:crates}. This section instead follows the control path
the paper's contribution rests on: what happens to one job between
resolution and the audit record.

\code{ox run} opens \code{.oxymake/state.db}, begins a run record, and
registers every job of the topological order against that record before
the scheduler starts. The scheduler then consults the cache for each
ready job through a single trait method, and the validation policy in
force decides how much of the path below it walks.

\paragraph{Key computation.}
\code{ox-cache} frames the seven components of
Equation~\ref{eq:cache-key}---key-format version, rule source, the
content hashes of the declared inputs, parameters, environment
specification, shell, and platform---into one BLAKE3 digest. Under
\code{--cache-validation=mtime} no key is computed \emph{on the lookup
path}; it is still computed at registration.

\paragraph{Lookup.}
Under \code{mtime} the lookup is a stateless filesystem comparison:
every declared output must exist and be newer than the newest declared
input, with no key and no database read. Under \code{mtime+hash} and
\code{hash} the key selects a row in \code{cache\_entries} and the
recorded \code{output\_records} for that key; if a directory remote
cache is configured, the outputs may first be restored from it by
content hash.

\paragraph{Validation.}
Each recorded output is then checked against the policy. A missing
output invalidates the entry outright. \code{mtime} compares
modification time and size only; \code{mtime+hash} trusts the recorded
content hash when metadata match and re-hashes the file when they do
not; \code{hash} re-hashes unconditionally. Any mismatch removes the
entry, and the job is treated as a miss.

\paragraph{Execution.}
A miss dispatches the job to the selected executor
(\S\ref{sec:distributed}); a hit marks it \code{Skipped} without running
it.

\paragraph{Registration.}
On completion, and only if every declared output exists on disk, the
key is recomputed together with its components and one SQLite
transaction upserts the cache entry---completion time, reproducibility
class, input hashes, job-specification hash---and replaces the output
rows, storing each output's content hash, modification time and size.
A configured remote cache then receives the output bytes under their
own content hash.

\paragraph{Audit commit.}
After the scheduler returns, the per-job durations collected from the
event stream are written to the history table and the run record is
closed with its succeeded, failed and skipped counts
(\S\ref{sec:state}).

\noindent The intended boundaries are that \code{ox-\allowbreak{}core}
makes no network calls; \code{ox-\allowbreak{}format} never
validates file existence or executes rules; and \code{ox-\allowbreak{}state} never
decides whether a job should re-run.  What Cargo enforces is narrower
than this: it prevents a crate from calling another \emph{workspace}
crate it does not depend on, but \code{std::fs}, \code{std::net} and
\code{std::process} are linked into every crate without appearing in any
manifest, and ``never decides whether a job should re-run'' is a
semantic property no dependency graph constrains. The remaining
boundaries are conventions maintained by review, and one of them is not
held today: \code{ox-\allowbreak{}core} does perform file I/O, in its
async disk writer (which stages output bytes to a per-attempt temporary
file, fsyncs, renames, then fsyncs the parent directory) and on the
scheduler's in-memory input path.

\subsection{Machine-Readable Execution Interface}
\label{sec:agent-api}

Every \code{ox} command supports \code{--json} for structured output.
When active, events are emitted as newline-delimited JSON (NDJSON) on
stdout:

\begin{lstlisting}[style=shell]
{"type":"run.started",
 "total_jobs":10001,
 "to_run":847,"cached":9154}
{"type":"job.completed",
 "id":"align_S001","duration_ms":272000}
{"type":"gate.reached",
 "id":"qc_check",
 "message":"Review QC metrics"}
\end{lstlisting}

A downstream tool reads line-by-line, reacting to typed events. Gate
decisions are commands; \code{ox gate list} prints pending gates with a
numeric identifier, which \code{approve} and \code{reject} take:

\begin{lstlisting}[style=shell]
ox gate approve 3 \
  --approver "ci:qc-runner" \
  --reason "metrics within threshold"
\end{lstlisting}

\paragraph{Enforcement status.}
In v0.1.0 the run path constructs the scheduler without a gate checker:
\code{[gate.*]} declarations are parsed, \code{ox gate} records and
lists decisions, and the scheduler's blocking logic and the
\code{gate.reached} event exist and are exercised by tests, but no
shipped command supplies the checker that would connect them. A rule
guarded by a gate therefore runs without approval, and \code{ox gate
list} reports no pending gate, because nothing registers one during a
run. Wiring a checker backed by the state database, and a \code{lint}
warning until it lands, are tracked as issue~\#2 of the repository. The
event stream and the approval commands described above are real; the
checkpoint they are meant to implement is not yet in force.

For Rust embedding, the scheduler exposes a typed
\code{tokio::\allowbreak{}broadcast} channel of \code{Event}
values, enabling in-process integration without stdout parsing.

\paragraph{MCP server.}
A subset of these operations is exposed over the Model Context Protocol
through \code{ox serve --mcp}, an stdio server (crate \code{ox-mcp}).
The tool catalogue is deliberately narrow: eight tools, of which seven are
read-only inspection---\code{ox\_status}, \code{ox\_plan}, \code{ox\_dag},
\code{ox\_logs}, \code{ox\_history}, \code{ox\_lint},
\code{ox\_explain}---plus one mutating tool, \code{ox\_clean}, which
removes cached artifacts and state. Notably there is \emph{no}
\code{ox\_run} tool and no gate-approval tool: an agent can plan,
inspect, and explain a workflow over MCP, but starting execution and
approving a gate remain human- or script-driven through the CLI. This is
a conservative default rather than a limitation of the protocol, and the
catalogue is pinned by a test so it cannot widen silently. The server is
an interface, not a durability layer: it adds no queueing, retry, or
cross-host recovery, and an agent driving external side effects must
supply its own idempotency and approval. We report the interface as
implemented and do not evaluate agent task performance over it.

\subsection{State Management}
\label{sec:state}

Three logically separated state concerns share a physical SQLite database\linebreak
(\code{.oxymake/\allowbreak{}state.db}):

\begin{description}
\item[Execution state] (ephemeral): What is running/pending/done.
  Reconstructible from the DAG and cache if lost. Uses WAL mode and
  atomic transactions for concurrent access.

\item[Cache state] (persistent, immutable): Content-addressable store
  keyed by Equation~\ref{eq:cache-key}. Directory structure:
  \code{.oxymake/cache/\{prefix\}/\{hash\}}. The artefact bytes are
  independent of SQLite and can be shared across same-platform machines
  through the directory remote cache (\code{ox run --cache-remote
  <dir>}), which verifies every restored artefact by content
  (index locality: \S\ref{sec:content-addressable}). S3/GCS transports
  are not implemented (\S\ref{sec:feature-matrix}).

\item[Audit state] (append-only): Run history with notes, per-job metrics
  (wall time, peak memory, exit code, hostname, environment hash). This
  provenance record enables reproducibility audits and performance
  regression detection. Metrics not recorded at the time of a run
  cannot be reconstructed afterwards; the table is append-only by
  convention, and the engine provides no tamper-evidence.
\end{description}

\paragraph{What does not travel.}
The three concerns differ in what a second checkout inherits. Artefact
bytes do: the directory remote cache stores them under their own content
hash, and any checkout can fetch them. The index does not. In v0.1.0 the
rows that map a computation key to its output paths and hashes stay
local to each checkout, and a remote restore is attempted only for a key
that checkout already holds an entry for---so a fresh checkout pointing
at the same shared directory re-executes the job rather than restoring
it. The audit trail is likewise per-workspace
(\S\ref{sec:content-addressable}, \S\ref{sec:limitations}).

Schema versioning uses \code{PRAGMA user\_version} with automatic
migrations from the first release; a migration runs inside a
transaction, so on local disk---where the
\code{StateDbAtomicCommit} axiom holds (\S\ref{sec:named-invariants})---a
\code{state.db} is never left half-migrated. On a shared filesystem
that premise fails and the property is not claimed
(\S\ref{sec:limitations}).

\section{Evaluation}
\label{sec:evaluation}

We evaluate OxyMake along eight axes: performance benchmarks on synthetic
workflows (Section~\ref{sec:perf-bench}), end-to-end execution
(Section~\ref{sec:e2e-bench}), the content-addressing
behaviour under mtime churn (Section~\ref{sec:git-checkout-bench}), a
scalability study with
algorithmic optimization (Section~\ref{sec:scale-study}), Snakemake
compatibility (Section~\ref{sec:snakemake-compat}), startup overhead
(Section~\ref{sec:startup-bench}), binary footprint
(Section~\ref{sec:binary-bench}), and FAIR compliance
(Section~\ref{sec:fair-eval}). All experiments were run on a single Apple
M4~Max (16-core, 128\,GB RAM, arm64); the head-to-head harness ran on
2026-08-18 under Darwin~25.5.0 and the startup and binary-footprint
micro-benchmarks on 2026-08-27 at commit \code{1a4f724} under
Darwin~25.5.0. No
Linux/x86\_64 measurements were taken, and no cross-architecture claim
is made. The DAG-resolution
and end-to-end head-to-head numbers (Tables~\ref{tab:dag-resolution}
and~\ref{tab:e2e}) come from a single head-to-head benchmark harness,
\code{bench/snakemake-vs-oxymake/} in the public repository: the
sub-second resolution phase is
timed with \code{hyperfine} (median over its statistical runs), and the
minutes-scale end-to-end phase with \code{/usr/bin/time} as a single
timed run per cell (the harness's \code{RUNS} setting governs the
resolution phase only). The measured build is the in-tree
\code{cargo build --release} binary at commit \code{03864f8} of the
public repository (invoked as \code{target/release/ox}, not a binary
resolved from \code{\$PATH}), against Snakemake~7.32.4 installed with
\code{pip}; OxyMake ran at \code{-j 16} and Snakemake at
\code{--cores 16} on an otherwise idle host, and every end-to-end row
names the OxyMake \code{--cache-validation} policy that produced it.
Dispersion (IQR or confidence intervals) is not
reported; the statistical limits of the evaluation are stated in
\S\ref{sec:limitations}. The startup and binary-footprint
micro-benchmarks were timed separately; their per-run samples are in
\code{docs/paper/experiment-results.md} in the public repository.

\subsection{DAG Resolution Performance}
\label{sec:perf-bench}

All numbers in this section come from a single benchmark harness in
the public repository,
\code{bench/\allowbreak snakemake-\allowbreak vs-\allowbreak oxymake/},
with a one-line reproducer (\code{bash run.sh} from that directory).
The harness drives a synthetic four-layer DAG
(\code{seed} $\to$ $N$ \code{gen} shell rules $\to$ $N$ \code{process}
Python-via-shell rules $\to$ $N$ \code{finalize} file-copy rules $\to$
\code{merge}), totalling $3N{+}2$ jobs; the three scales use $N{=}33$,
$333$, and $3333$, i.e.\ exactly 101, 1{,}001, and 10{,}001 jobs, and
the tables label them by these exact counts. The same DAG and per-job work are declared once
as \code{workflow.toml} (OxyMake) and once as \code{workflow.smk}
(Snakemake~7.32.4); only the orchestrator changes between runs. We
benchmark the Snakemake~7.x line because it is the version that added
recorded provenance to the mtime comparison, and is therefore the
strongest change-detection comparator; the 8.x line changes the executor
and deployment model but not the provenance-based rebuild decision this
paper compares against. DAG resolution
time was measured using \code{ox plan} for OxyMake and
\code{snakemake --dryrun} for Snakemake---both parse the workflow, resolve all
wildcards, build the job graph, and report the execution plan without running
any jobs. The resolution phase is millisecond-scale
(5.9--118.3\,ms across the measured scales), so we time it with
\code{hyperfine} (statistical warmup, no subprocess-wrapper overhead); the
end-to-end phase below is minutes-scale and timed with \code{/usr/bin/time}.
One scope note applies to every head-to-head number in this section
and the next, naming exactly which validation mode produced it. DAG
resolution is not symmetric on this point: \code{ox plan} exercises no
cache validation, whereas building the DAG in \code{snakemake --dryrun}
evaluates the rerun triggers for every job, so the dry run stats inputs
and outputs and reads \code{.snakemake/metadata}. No input checksum is
computed on either measured row: on the cold row a missing output
already supplies the rebuild reason, and on the warm row every input
predates the outputs it feeds, so the checksum comparison is
short-circuited (\code{dag.py}, v7.32.4). The ratio therefore bounds rather than
isolates the resolution difference. The cold and warm end-to-end rows run
Snakemake under its default rerun-triggers (recorded per-output
provenance) at \code{--cores 16}, and OxyMake at \code{-j 16} under
the shipped \code{--cache-validation=mtime+hash} default
(\S\ref{sec:content-addressable}); two additional warm runs pin
\code{--cache-validation} to \code{mtime}, the metadata fast path, and
to \code{hash}, full content re-verification. All three warm policies
are measured separately and every row below names the one that
produced it. The
measurements that back the content-addressing thesis are that
\code{hash}-mode row and the \code{git checkout} scenario---one shared
input's mtime perturbed, no content changed, count the jobs each engine
re-runs---reported in \S\ref{sec:git-checkout-bench}.

\begin{table}[htbp]
\centering\small
\begin{tabular}{@{}rrrr@{}}
\toprule
\textbf{Jobs} & \textbf{OxyMake \code{ox plan}} & \textbf{Snakemake \code{--dryrun}} & \textbf{Speedup} \\
\midrule
101      &    5.9\,ms  &    595.1\,ms  & 100.86$\times$ \\
1{,}001  &   13.1\,ms  &    792.4\,ms  & 60.49$\times$ \\
10{,}001 &  118.3\,ms  & 2{,}721.7\,ms  & 23.01$\times$ \\
\bottomrule
\end{tabular}
\caption{DAG-resolution wall-clock time, median over hyperfine runs
  (benchmark harness \code{bench/snakemake-vs-oxymake/}). Measured 2026-08-18,
  Apple M4~Max (Darwin~25.5.0, 16 cores); OxyMake \code{ox plan},
  Snakemake~7.32.4 \code{snakemake --dryrun}. Times are shown at the
  harness's recorded precision (0.1\,ms), so each speedup is the
  quotient of the two times printed in its row. Raw
  measurements and the reproducer are in
  \code{bench/snakemake-vs-oxymake/RESULTS.md} in the public repository.}
\label{tab:dag-resolution}
\end{table}

\noindent OxyMake resolves a 10{,}001-job DAG in 118.3\,ms versus
Snakemake's 2{,}721.7\,ms, a 23.01$\times$ difference on DAG resolution. The ratio
\emph{narrows} with scale (100.86$\times$ at 101 jobs to 23.01$\times$ at 10{,}001),
which is consistent with a fixed per-invocation cost in the Snakemake
run being amortised over more jobs while OxyMake's resolution grows
with job count. A hundredfold increase in job count, from 101 to
1{,}001 to 10{,}001, moves \code{snakemake --dryrun} only from 595.1\,ms
to 792.4\,ms to 2{,}721.7\,ms, so the 101-job row is dominated by cost that
does not scale with the graph. These aggregate timings do not allocate
that fixed cost among interpreter startup, imports, metadata work and
resolution---we ran no phase-level profile---so the
100.86$\times$ entry should not be read as a resolution-only ratio. The
marginal ratio at 10{,}001 jobs---derived from a linear fit through the
two endpoints---is $\approx$19$\times$. From 101 to 10{,}001 jobs ($\approx$100$\times$
more work) OxyMake adds $\approx$112\,ms, a marginal cost of $\approx$11\,$\mu$s/job,
consistent with a per-target lookup whose cost is a handful of precompiled
regex matches at this rule count (Section~\ref{sec:scale-study}).
\textbf{This comparison is scoped to DAG resolution}; the end-to-end
comparison, in which OxyMake is slower on a cold run, is reported next.

\subsection{End-to-End Execution}
\label{sec:e2e-bench}

DAG resolution is the phase OxyMake optimises; it is not the whole story. The
same harness also measures \emph{end-to-end} wall time---resolve plus execute
every job---and here OxyMake is \emph{slower} than Snakemake on a cold run.

\begin{table}[htbp]
\centering\small
\begin{tabular}{@{}rrrr@{}}
\toprule
\textbf{Jobs} & \textbf{Snakemake} & \textbf{OxyMake} & \textbf{OxyMake / Snakemake} \\
\midrule
101      &   1.623\,s  &    2.386\,s  & 1.47$\times$ (slower) \\
1{,}001  &   6.747\,s  &   17.367\,s  & 2.57$\times$ (slower) \\
10{,}001 &  92.666\,s  &  196.934\,s  & 2.13$\times$ (slower) \\
\bottomrule
\end{tabular}
\caption{Cold end-to-end wall time (same benchmark harness, measured
  2026-08-18; OxyMake \code{-j 16}, Snakemake \code{--cores 16}).
  OxyMake runs
  $1.47$--$2.57\times$ slower than Snakemake on a cold execution across scales.
  Times are shown at the harness's recorded precision (1\,ms), so each
  ratio is the quotient of the two times printed in its row. The
  benchmark's rendered summary (\code{RESULTS.md}) prints the
  10{,}001-job row rounded to minutes scale;
  the seconds here are the same samples unrounded.
  Warm-cache re-run is the inverse (next paragraph).}
\label{tab:e2e}
\end{table}

\noindent On a first, cold execution both engines record per-job
provenance: Snakemake~7.32.4 writes rule code, parameters, the input
set and the software environment into \code{.snakemake/metadata} at
every job completion, together with a SHA-256 checksum of each eligible
input. The work OxyMake performs that Snakemake does not is to fold the
rule source, input content hashes, parameters, environment
specification and platform into a single BLAKE3 cache key, and to hash
every \emph{output} and record its content hash in the local cache
index. (No \code{ox.lock} is written by \code{ox run}, and no artefact
bytes are copied into a blob store on this workload: the shared blob
store is opt-in via \code{--cache-remote}, which the harness does not
set.) A first cold run has no prior
provenance for either engine to consult. We attribute the measured
cold differential to this key computation and to the output hashing and
index writes; this
attribution is a hypothesis consistent with the design, not the result
of a phase-level profile, which we have not performed. The overhead is
real wall-clock cost on the cold path,
exchanged for content-addressable correctness, cache-key validity
across same-platform machines, and an auditable rebuild decision
(\S\ref{sec:fair-eval}). Three measurements on the \emph{same} harness
show what that bookkeeping provides:

\begin{itemize}
  \item \textbf{Warm-cache re-run} (the common inner-loop case): at 10{,}001 jobs,
    with both engines at 16-way parallelism, OxyMake completes a no-op
    rebuild in 1.378\,s under the shipped \code{mtime+hash} default
    versus Snakemake's 3.478\,s---2.52$\times$ faster. Pinning
    \code{--cache-validation=mtime}, the metadata fast path, brings the
    same no-op rebuild to 500.1\,ms---6.95$\times$ faster---and
    \code{--cache-validation=hash}, which re-hashes every input and
    output before deciding, to 1.821\,s---1.91$\times$ faster, with
    every byte verified. The default sits between the two: it computes the
    cache key and consults the index, which the \code{mtime} fast path
    skips, but re-hashes only files whose metadata moved, which
    \code{hash} does not skip (\S\ref{sec:content-addressable}).
  \item \textbf{Rebuild scope}: rewriting the content of one
    Layer-1 input, the expected re-run scope is 3 jobs
    (\code{process}, \code{finalize}, \code{merge}). Snakemake re-runs
    exactly 3 at every scale; OxyMake re-runs 4, one more than the
    minimum, at every scale. OxyMake's content-addressed decision is
    therefore tight but not measured to be minimal on this workload;
    the extra job is a scope overshoot we have not diagnosed. Its
    computation key is identical on any machine of the
    same platform where a per-tree record is bound to one tree (the
    platform term in Eq.~\ref{eq:cache-key} scopes the key to one
    OS/arch pair; heterogeneous reuse is future work, and the
    rebuild-decision preconditions are stated in
    \S\ref{sec:introduction}).
  \item \textbf{Memory}: peak resident set of the orchestrator on the cold
    10{,}001-job run is 89.9\,MiB for OxyMake versus 184.5\,MiB for
    Snakemake---2.05$\times$ smaller.
\end{itemize}

\noindent In summary, on this workload OxyMake is 1.47--2.57$\times$
slower on cold runs and, on warm no-op runs at 10{,}001 jobs,
2.52$\times$ faster under the shipped \code{mtime+hash} default,
6.95$\times$ under \code{mtime} and 1.91$\times$ under \code{hash}. These results should not be
generalized beyond the evaluated workload.
Section~\ref{sec:scale-study} treats DAG-resolution scaling; the FAIR
contract that the cold-path bookkeeping underwrites is evaluated in
Section~\ref{sec:fair-eval}.

\subsection{Content-Addressing under mtime Churn: the git-checkout Test}
\label{sec:git-checkout-bench}

The scenario that separates timestamp-trusting from content-checking
validation is mtime churn: a shared input's timestamp moves, no byte
changes. The
harness reproduces it by bumping the mtime of a shared tracked input
(\code{bench\_lib.py}, a declared \code{lib} input of every
\code{process} job) after a clean build. This models the mtime-only
perturbation that a \code{git checkout}, a tree copy, or a
backup-restore introduces: metadata moves, content does not. Those
operations differ in other respects; the benchmark isolates the mtime
change they share. A decision that trusts timestamps must re-run the
$N$ \code{process} jobs that read the file and, by the downstream
cascade, their $N$ \code{finalize} dependents plus \code{merge}---$2N{+}1$
in total; a decision that checks content must re-run zero.

\begin{table}[htbp]
\centering\small
\begin{tabular}{@{}rrrr@{}}
\toprule
\textbf{Jobs} & \textbf{Snakemake 7.32.4} & \textbf{OxyMake \code{mtime}} & \textbf{OxyMake \code{hash}} \\
\midrule
101      & 0 &      67 & 0 \\
1{,}001  & 0 &     667 & 0 \\
10{,}001 & 0 & 6{,}667 & 0 \\
\bottomrule
\end{tabular}
\caption{Jobs re-run after bumping a shared input's mtime without
  changing a byte (same benchmark harness, measured 2026-08-18).
  Validation modes are explicit: Snakemake runs its default
  rerun-triggers (recorded provenance); the OxyMake columns pin
  \code{--cache-validation} to \code{mtime} and \code{hash}
  respectively. The shipping \code{mtime+hash} default is not a
  separate measured column; its zero re-run count follows from the
  policy definition (\S\ref{sec:content-addressable}), and what the two
  measured modes bracket is its warm-run \emph{time} (see text).}
\label{tab:mtime-churn}
\end{table}

\noindent\textbf{Pure \code{mtime} validation re-runs the full radius.} It re-runs
$2N{+}1$ jobs at every scale (6,667 at the largest): the metadata-only
policy re-runs the $N$ jobs that read the perturbed file, and the
cascade carries their $N$ dependents and \code{merge} with them. The
\code{mtime+hash} policy re-hashes a file whose metadata moved before
deciding, so a churned tree re-runs zero jobs; the only cost is the
re-hash of the perturbed files. Its re-run count in this scenario was
not measured as a separate column, but its warm no-op \emph{time} was:
at 10{,}001 jobs it is 1.38\,s, between the 500\,ms of the
metadata-only path and the 1.82\,s of re-hashing everything
(\S\ref{sec:e2e-bench}).

\textbf{Snakemake~7.32.4 re-runs zero jobs.} Under its default
rerun-triggers it re-ran zero jobs, both under a plain \code{touch} and
with a far-future mtime. The mechanism is not that live
timestamps are ignored---they are compared first---but that an input
newer than the oldest output counts as updated only if its recorded
SHA-256 also fails to match. The perturbed file
(\code{bench/snakemake-vs-oxymake/bench\_lib.py}) is 638 bytes, below the
100{,}000-byte threshold above which Snakemake records no checksum, so
the checksum decides here; a larger shared input would leave the mtime
comparison to decide alone. The phantom-re-run failure mode is real
for GNU Make and for any engine's pure-mtime fast path, but the
benchmarked Snakemake version does not exhibit it.

\textbf{Content-addressing does not improve this scenario on one
machine.} On a single working tree, \code{hash} mode reaches parity
with Snakemake's provenance, not superiority. What a
working-tree-bound provenance record cannot offer is: a
computation key that is a pure function of content and workflow-root-relative
paths, hence an identity that is identical on any same-platform
machine (\S\ref{sec:content-addressable}); detection
of on-disk output corruption that a provenance record would serve
stale (\S\ref{sec:content-addressable}); and a plan and key derivation
auditable from the \code{ox.lock} record, the rebuild decision itself
being auditable given the checkout's live reuse state---local
key-to-output index, materialised outputs, and validation policy
(\S\ref{sec:introduction}, \S\ref{sec:fair-eval}).
It also avoids the pure-\code{mtime} re-run radius the
middle column documents.

\subsection{Scalability: ProducerIndex Optimization}
\label{sec:scale-study}

The measured resolution times in Table~\ref{tab:dag-resolution}
($\leq$118\,ms at 10{,}001 jobs) are already small, but the naive
backward-chaining algorithm parses and compiles every candidate output
pattern for every target it resolves---$O(T \times P)$ pattern
compilations to resolve $T$ targets against $P$ output patterns, with a
large constant we attribute to regex construction (a design argument; we
ran no phase-level profile).
We implemented a \code{ProducerIndex}: all output patterns are parsed
and compiled \emph{once}, up front, so each target lookup is a scan of
precompiled regexes rather than a parse-and-compile cycle.  The
asymptotic complexity is then a one-time $O(P)$ compilation plus
$O(T \times P)$ lookup---each lookup is still a
linear scan over the $P$ compiled patterns---but the constant is
expected to drop, because per-target regex compilation is the term the
index removes.  We did not measure the pre-index implementation, so the
improvement is a design argument, not a measured speedup.  A hash-based prefix index that would give expected $O(P)$
construction and $O(T)$ lookup is future work, not part of the measured
system.  The scaling observed across
$101 \to 10{,}001$ jobs in Table~\ref{tab:dag-resolution} (5.9\,ms~$\to$~13.1\,ms~$\to$~118.3\,ms:
$\approx$112\,ms added for a $\approx$100$\times$ increase in job count, $\approx$11\,$\mu$s/job)
is near-linear in job count because the rule count---and hence the
per-target scan cost---stays small and fixed while the job count grows;
quadratic behaviour would surface only if rules and targets grew
together.  Beyond the measured range we report only projections: extrapolating
the slope suggests $\sim$0.6\,s at $5\times10^4$ jobs and $\sim$1.1\,s at $10^5$
jobs. \textbf{These are algorithmic-headroom projections, not measurements}---the
harness was run to 10{,}001 jobs only; $5\times10^4$ and $10^5$
were not measured.

\subsection{Workflow Language Compatibility}
\label{sec:snakemake-compat}

The \code{ox translate} command accepts both Snakemake and WDL workflows
and converts them to Oxymakefile TOML, while \code{ox export snakemake}
and \code{ox export wdl}
convert an Oxymakefile back to the respective formats. Source format is
auto-detected from file extensions (\code{.smk}, \code{.wdl}) or content
inspection. \textbf{Only the Snakemake import direction is evaluated
here}: the export direction and WDL translation are described but not
exercised by any experiment in this paper.

\paragraph{Snakemake.}
We exercised Snakemake translation on four workflows of increasing
complexity: an eight-rule tutorial pipeline over three samples, a
six-rule RNA-seq counting workflow over four samples driven by a
\code{configfile}, a six-rule CSV ETL pipeline over three sources, and a
five-rule multi-sample QC workflow over five samples. The workflows, their
translated Oxymakefiles, and the full result log are in the public
repository under \code{benchmark/snakemake-compat/} (see
\code{RESULTS.md} there). The recorded outcome is per-workflow
translation, dry-run and execution status together with the number of
manual fixes; the record contains no source-versus-translated comparison
of the output file tree, so no output-equivalence result is claimed
here.

\textbf{None of the four ran unmodified.} All four executed to completion
only \emph{after} manual fixes to the translated Oxymakefile, between
three and five edits per workflow. The recurring interventions were:
adding \code{expand = "product"} to rules that aggregate over wildcards
(the translator escalates \code{expand()} calls that reference a
Python-scoped variable instead of resolving them); replacing
\code{\{input.name\}} with unnamed \code{\{input\}} in aggregation rules,
because a named input under \code{expand} substitutes only the first
matching path; rewriting Snakemake's \code{\{\{\}\}} brace escaping;
replacing bash-only shell constructs, since OxyMake executes under
\code{/bin/sh}; and naming targets explicitly, because rules are stored in
a \code{BTreeMap} so the default target is the alphabetically first rule
rather than \code{rule all}. Two of the four translated with no escalation
at all and still needed four hand edits each, so a clean translation
report is not evidence of a runnable workflow.

Translation handles rule definitions, wildcard expansion,
\code{params}/\code{threads}/\code{resources} blocks,
\code{wildcard\_constraints}, and \code{configfile} references.
Unsupported Snakemake features (embedded Python expressions, \code{run:}
blocks with arbitrary code, dynamic thread counts, \code{wrapper},
\code{shadow}/\code{group}/\code{envmodules}) emit structured warnings or
escalations with migration guidance. The honest summary is that
\code{ox translate} is a migration aid that removes most of the mechanical
work, not a drop-in transpiler.

\paragraph{WDL (Workflow Description Language).}
WDL~\cite{vossWDLCromwell2017} is the standard workflow language in genomics,
used with Cromwell, miniWDL, and the Terra platform. OxyMake's WDL translation
maps WDL's \code{task}/\code{workflow}/\code{call} model to OxyMake's rule-based
model: each WDL task becomes an OxyMake rule, \code{command} blocks become
\code{shell} directives, and \code{runtime} blocks (docker, cpu, memory, disks)
map directly to OxyMake's resource and environment declarations.  WDL's
\code{scatter} blocks are translated to OxyMake's expand mode, and WDL placeholder
syntax (\verb|~{var}| and \verb|${var}|) is converted to OxyMake's
\verb|{var}| format.  WDL's richer type system (\code{File}, \code{Array[File]},
\code{Int}) produces structured escalations for constructs that require manual
review, such as optional inputs (\code{File?}) and glob expressions.

\paragraph{Why TOML, not WDL?}
WDL is optimized for a specific domain: bioinformatics pipelines running on
cloud infrastructure via engines like Cromwell. Its typed, portable
specification is well suited to genomics workflows but requires
declarations that general computational pipelines do not need: every
input and
output must carry an explicit type declaration, and the
workflow/call/scatter hierarchy adds structural overhead for simple
file-to-file transformations. (The \code{runtime} section and its
individual attributes are \emph{optional} in WDL~1.2.) OxyMake's TOML
format is domain-agnostic---
a three-line rule with \code{input}, \code{output}, and \code{shell} suffices
for simple cases, while the same format scales to complex multi-wildcard
pipelines with resource declarations.  OxyMake ships as one \code{ox}
executable requiring no OxyMake daemon for local execution; WDL requires a separate
execution engine (Cromwell, miniWDL). The translator is intended to bridge
both ecosystems in both directions, but the WDL paths are syntactic
mappings whose adequacy this paper does not evaluate: no WDL workflow
was translated, executed, or round-tripped here.

\subsection{Startup Time}
\label{sec:startup-bench}

We measured cold-start time for three commands of increasing complexity:

\begin{table}[htbp]
\centering\small
\begin{tabular}{@{}lr@{}}
\toprule
\textbf{Command} & \textbf{Wall-clock time} \\
\midrule
\code{ox --help}             & 2.2\,ms \\
\code{ox lint} (10 rules)    & 2.4\,ms \\
\code{ox lint} (1{,}000 rules) & 5.9\,ms \\
\bottomrule
\end{tabular}
\caption{Startup time (\code{hyperfine} median of 200~runs, shell
  bypassed; measured 2026-08-27 at commit \code{1a4f724} on the same
  host under Darwin~25.5.0 and recorded in
  \code{docs/paper/experiment-results.md}).}
\label{tab:startup}
\end{table}

\noindent Startup time is 2.2--5.9\,ms across the three commands
measured. An earlier run of these three commands (2026-04-01, on a
52{,}514-SLOC tree) reported 6--7\,ms under a Python
\code{subprocess} harness whose spawn overhead is included in the
figure; the numbers above were re-measured at the evaluated build with
\code{hyperfine -N}, so they are not directly comparable with it. The
native Rust binary carries no Python or JVM startup overhead. These
numbers are a component measurement rather than a usability comparison.

\subsection{Binary Size}
\label{sec:binary-bench}

\begin{table}[htbp]
\centering\small
\begin{tabular}{@{}lr@{}}
\toprule
\textbf{Metric} & \textbf{Value} \\
\midrule
Binary size (release, default features) & 17.4\,MB \\
Binary type & Mach-O 64-bit arm64 \\
Full release build time (clean tree) & 26\,s \\
\bottomrule
\end{tabular}
\caption{Binary footprint and build metrics (measured 2026-08-27 at
  commit \code{1a4f724} on the same host under Darwin~25.5.0; the
  binary is 17{,}421{,}504 bytes, and the build time is the wall clock
  of \code{cargo build --release} after \code{cargo clean --release}
  with the dependency sources already fetched. Recorded in
  \code{docs/paper/experiment-results.md}).}
\label{tab:binary}
\end{table}

\noindent The 17.4\,MB binary includes all default-feature crates
(among them the TUI monitor, web dashboard, and Snakemake/WDL translator). It is
one self-contained executable that bundles no interpreter and needs no
OxyMake daemon, linking only the host platform's system libraries;
installable via \code{cargo install} or direct download.

\subsection{Implementation Status}
\label{sec:feature-matrix}

Table~\ref{tab:features} is an implementation matrix: it lists the
features that are implemented and marks which are exercised by the
benchmark harness. Implemented and evaluated are distinct: the Slurm
and Ray executors, the MCP server, the dashboard, and the directory
remote cache are implemented but not measured in this evaluation
(\S\ref{sec:evaluation}).

\begin{table}[htbp]
\centering\small
\begin{tabular}{@{}lp{0.55\linewidth}@{}}
\toprule
\textbf{Feature} & \textbf{Notes} \\
\midrule
TOML-based workflow definition   & Oxymakefile.toml parsing \\
Backward-chaining DAG resolution & Wildcards, fan-out \\
Content-addressable caching      & BLAKE3 + mtime fast-path \\
Three-graph architecture         & Rule $\to$ Job $\to$ Exec \\
Cache-pruning optimization pass  & Skips up-to-date jobs \\
Four execution modes             & shell, run, script, call \\
NDJSON event API (\code{--json}) & All commands \\
Gate approval                    & \code{ox gate approve/reject}; declarations parsed and decisions recorded, not yet enforced by the scheduler (\S\ref{sec:agent-api}) \\
Idempotent convergent execution  & SQLite state; claim protocol implemented but not yet the scheduling gate (\S\ref{sec:daemon-free}) \\
Tags and \code{--where} filter & Implicit + explicit tags \\
Conditional guards (\code{when}) & DAG-time evaluation \\
Lockfile (\code{ox.lock})      & Re-derivable plan of record \\
Snakemake translator             & \code{ox translate}, \code{ox export snakemake} \\
WDL translator                   & \code{ox translate}, \code{ox export wdl} \\
TUI monitor (\code{ox top})    & Real-time dashboard \\
Web dashboard                    & HTML status page \\
Slurm executor                   & \code{--executor slurm} \\
MCP server                       & Model Context Protocol endpoint \\
Ray executor                     & \code{--executor ray} \\
Directory remote cache           & \code{--cache-remote}, blob store (\S\ref{sec:content-addressable}) \\
\bottomrule
\end{tabular}
\caption{Implementation matrix. All rows are implemented, with the two exceptions the notes column states (gate enforcement and the claim gate); the benchmark
  harness exercises TOML parsing, DAG resolution, content-addressable
  caching, cache pruning, and the \code{shell} execution mode on the
  local executor (its Python and file-copy work is launched through
  shell rules). The \code{run}, \code{script}, and \code{call} modes
  are covered by the test suite but not by the benchmark. The remaining
  rows are implemented but not measured here.}
\label{tab:features}
\end{table}

\noindent Kubernetes execution, object-store (S3/GCS) cache backends,
in-memory data passing on the local executor, and five of the six
optimization passes are not implemented and are excluded from the
evaluation.

\subsection{FAIR Compliance Assessment}
\label{sec:fair-eval}

We assess OxyMake against the FAIR principles for workflows argued for
by Goble et al.~\cite{gobleFAIRComputationalWorkflows2020} and made
concrete as numbered entries by Wilkinson et
al.~\cite{wilkinsonApplyingFAIRPrinciples2025}, and against
the FAIR4RS principles for the engine
itself~\cite{chueHongFAIR4RSPrinciples2022}.
Table~\ref{tab:fair} is a self-assessment: each row records the mechanism
that supports the indicator, not an external audit or a third-party
certification. It mixes properties of the engine, the workflow format,
and the produced artefacts, and a reader should read each cell as ``the
mechanism is present'', not ``the indicator is independently verified''.

\begin{table}[htbp]
\centering\footnotesize\setlength{\tabcolsep}{4pt}
\begin{tabular}{@{}lllp{0.40\linewidth}@{}}
\toprule
\textbf{Principle} & \textbf{Aspect} & \textbf{OxyMake} & \textbf{Mechanism} \\
\midrule
\multirow{3}{*}{Findable}
  & Unique ID        & Partial & Lockfile content hash; not a registry-backed persistent identifier \\
  & Rich metadata    & Partial & Machine-readable TOML; no standard metadata vocabulary \\
  & Searchable registry & Future & Workflow registry planned \\
\midrule
\multirow{2}{*}{Accessible}
  & Standard protocol & Partial & Files retrievable via Git/HTTP; artefact accessibility not guaranteed \\
  & Open format       & Native & TOML, no vendor lock-in \\
\midrule
\multirow{3}{*}{Interoperable}
  & Standard serialization & Partial & Data artefacts in Parquet/CSV/IPC; workflow language engine-specific \\
  & Workflow language       & Partial & TOML + WDL/Snakemake bridge \\
  & Cross-platform          & Not assessed & Compiles for three OSes; only macOS/arm64 exercised here \\
\midrule
\multirow{3}{*}{Reusable}
  & Provenance        & Partial & Audit trail in state.db: an engine-specific trace in a project-local file, with no standard provenance vocabulary and no portable archival package \\
  & Reproducibility   & Partial & Lockfile + content cache, witnessing declared-input identity and local reuse under the stated assumptions (\S\ref{sec:undeclared-inputs}; index locality, \S\ref{sec:content-addressable}) \\
  & Community standards & Partial & Open licence; no community packaging standard (e.g.\ RO-Crate) \\
\bottomrule
\end{tabular}
\caption{FAIR self-assessment. ``Native'' = the aspect is supported by
  a built-in mechanism; ``Partial'' = a mechanism exists but does not
  fully satisfy the aspect as defined by the cited papers;
  ``Not assessed'' = no evidence either way in this evaluation;
  ``Future'' = not yet implemented. The aspect labels are this paper's
  own condensation of the four principles; they are not the numbered
  indicators of Wilkinson et al.~\cite{wilkinsonApplyingFAIRPrinciples2025},
  whose Table~1 defines a different and finer set (F1--F4, A1--A2,
  I1--I4, R1--R3), and the rows should not be read as a row-by-row
  assessment against it.}
\label{tab:fair}
\end{table}

\noindent Under this self-assessment OxyMake supports one aspect
natively (open format), eight partially, one is not assessed
(cross-platform), and one is future work (searchable registry). The
largest gaps are the absence of a workflow
registry for discovery and the fact that the TOML workflow format
is not a community
standard like CWL. \code{ox translate} and \code{ox export} are
intended as bridges to the broader bioinformatics and genomics
ecosystems, but only the Snakemake import direction is exercised here
(\S\ref{sec:snakemake-compat}).

\subsection{Performance Summary}

\begin{table}[htbp]
\centering\small
\begin{tabular}{@{}llrrr@{}}
\toprule
\textbf{Scenario} & \textbf{Policy} & \textbf{OxyMake} & \textbf{Snakemake} & \textbf{Ratio} \\
\midrule
Cold end-to-end          & \code{mtime+hash} & 196.934\,s & 92.666\,s   & 2.13$\times$ slower \\
Warm no-op re-run        & \code{mtime+hash} &   1.378\,s &  3.478\,s   & 2.52$\times$ faster \\
Warm no-op re-run        & \code{mtime}      &  500.1\,ms &  3.478\,s   & 6.95$\times$ faster \\
Warm no-op re-run        & \code{hash}       &   1.821\,s &  3.478\,s   & 1.91$\times$ faster \\
mtime churn, jobs re-run & \code{mtime}      &  6{,}667   &  0          & --- \\
mtime churn, jobs re-run & \code{hash}       &  0         &  0          & --- \\
Peak RSS, cold run       & \code{mtime+hash} &  89.9\,MiB & 184.5\,MiB  & 2.05$\times$ smaller \\
DAG resolution           & ---               &  118.3\,ms & 2{,}721.7\,ms & 23.01$\times$ faster \\
\bottomrule
\end{tabular}
\caption{Head-to-head results on the benchmark harness's synthetic
  four-layer DAG at 10{,}001 jobs, both engines at 16-way parallelism
  (\S\ref{sec:perf-bench}--\S\ref{sec:git-checkout-bench}). The
  comparator is Snakemake~7.32.4 under its default rerun-triggers; the
  \textbf{Policy} column names the OxyMake \code{--cache-validation}
  setting that produced the row. The resolution row compares \code{ox
  plan} with \code{snakemake --dryrun}, which is not a symmetric
  comparison (\S\ref{sec:perf-bench}). The \code{mtime+hash} re-run
  count under mtime churn was not measured as a separate column
  (Table~\ref{tab:mtime-churn}). Startup time (2.2\,ms), binary size
  (17.4\,MB) and the \metricTests-test suite are separate
  micro-benchmarks, reported in
  \S\ref{sec:startup-bench}--\S\ref{sec:binary-bench}.}
\label{tab:perf-summary}
\end{table}

\section{Discussion}
\label{sec:discussion}

\subsection{Positioning in the Landscape}

Table~\ref{tab:landscape} places OxyMake against neighbouring systems on
four axes: the authoring surface, how the rebuild decision is made, how
the system is deployed, and whether it exposes a machine-readable
interface. No listed system combines all four properties
OxyMake combines---a file-rule surface, a
content-derived computation key as the default identity, a single daemon-free binary, and a
structured machine interface---though each property individually
appears in some neighbour.

\begin{table}[htbp]
\centering\footnotesize\setlength{\tabcolsep}{4pt}
\begin{tabular}{@{}lllll@{}}
\toprule
\textbf{System} & \textbf{Surface} & \textbf{Rebuild decision} & \textbf{Deployment} & \textbf{Machine interface} \\
\midrule
GNU Make            & file rules       & live mtime            & binary            & no \\
Snakemake~7         & file rules (Py)  & per-tree provenance   & Python runtime    & partial \\
DVC                 & pipeline stages  & content hashes        & Python runtime    & partial \\
Nextflow            & channels (Groovy)& task hash$^{\ddagger}$  & JVM runtime       & partial \\
cwltool + Arvados   & CWL standard     & content-sensitive cache & platform        & varies \\
Bazel / Buck2       & build targets    & content-addressed     & daemon (default)  & partial \\
Airflow             & DAG-as-code      & none (task re-runs)   & service           & yes \\
Prefect             & DAG-as-code      & none by default$^{\S}$ & service           & yes \\
Dagster             & asset graph (code)& data-version memoisation$^{\S}$ & service      & yes \\
Dask / Ray          & task/actor API   & (execution substrate) & runtime/cluster   & API \\
\textbf{OxyMake}    & file rules (TOML)& content key + index   & single Rust binary & yes (NDJSON, MCP) \\
\bottomrule
\end{tabular}
\caption{Positioning on four axes. The rebuild-decision column names
  what each engine's default decision reads; for OxyMake the
  content-derived key is the checkout-independent identity of a
  computation, and the decision additionally consults the materialised
  outputs, the validation policy, and---under the \code{mtime+hash}
  default and \code{hash}, but not under the opt-in \code{mtime}
  policy---the local key-to-output index
  (\S\ref{sec:content-addressable}). Slurm is not a row: it is an
  execution substrate OxyMake targets (\code{--executor slurm}), not a
  competitor. Dask and Ray are likewise substrates the executor
  backends consume rather than engines OxyMake replaces. Snakemake's
  opt-in \code{--cache} adds a content-keyed between-workflow cache
  beside the per-tree record without changing the default rebuild
  decision (\S\ref{sec:file-based}); DVC versions pipeline stages
  rather than expanding wildcard file rules (\S\ref{sec:file-based}).
  $^{\ddagger}$Nextflow's task hash covers the task script and its
  referenced variables, the task inputs, the container, environment
  modules and Conda/Spack environment, and the session; in the default
  caching mode file inputs enter it by path, size and last-modified
  time~\cite{nextflowCacheResume}.
  $^{\S}$Prefect does not persist results by default
  (\code{results.persist\_by\_default} is false), so it has no default
  cross-run reuse decision; its opt-in task cache keys on
  \code{INPUTS + TASK\_SOURCE + RUN\_ID}, whose run-ID component
  normally confines a hit to one parent run~\cite{prefectCaching}.
  Dagster derives an asset's data version by hashing its code version
  with the data versions of its inputs, reports the result as staleness,
  and can use those versions for a limited form of asset memoisation,
  skipping redundant materialisations and returning the previously
  computed value~\cite{dagsterVersioning}.}
\label{tab:landscape}
\end{table}

Compared
to Snakemake, OxyMake preserves the backward-chaining DAG paradigm while
replacing its rebuild decision---live mtimes in the Make tradition, or
working-tree-bound provenance records in Snakemake~7, which are recorded
rather than heuristic---with
content-addressable caching
(including transitive invalidation of deleted intermediates;
Section~\ref{sec:idempotent}),
the Python DSL with declarative TOML, and the Python runtime with a Rust
engine. Compared to Guix-based pipeline stacks---guix-cwl~\cite{prinsGuixCWLPipelines2018}
and the Guix-packaged Snakemake pipelines of
PiGx~\cite{wurmusPiGxReproducibleGenomics2018}---
OxyMake targets the orchestration layers (L2--L4) of the FAIR ladder
(Table~\ref{tab:fair-ladder}) and leaves the binary substrate (L1) to a
functional package manager; whether the two contracts compose
cleanly is a future direction (\S\ref{sec:future-work}), not a claim
made here.
Compared to service-deployment orchestrators,
OxyMake retains the simplicity of file-based rules while adding
content-addressable caching and a machine-readable API. Compared to build systems
(Bazel, Buck), the difference is semantics per axis rather than
presence versus absence of whole features: Bazel has target patterns,
configurable toolchains and platform constraints, and Starlark
extension points for policy, so it would be wrong to say it lacks
wildcards, environment management, or gates. What differs is what each
maps onto---OxyMake's wildcards expand \emph{output filenames} into
per-path jobs rather than selecting build targets; its environment
declarations name a scientific runtime (conda YAML, \code{uv.lock}, an
image) hashed by content into the job key rather than a toolchain
resolved by constraint; and its gates are declared as run-time human or agent
approval checkpoints inside a workflow rather than analysis-time policy
checks (declared and managed in v0.1.0, not yet enforced;
\S\ref{sec:agent-api}).

In the Mokhov et al.\ taxonomy~\cite[\S4--5]{mokhovBuildSystemsCarte2020}, OxyMake
combines a \emph{topological scheduler} (task order fixed before execution
from the dependency graph) with \emph{verifying traces} (skip a task when
its recorded input/output hashes still match disk) augmented by
content-addressing, with the deleted-output cascade as the dirty-bit
exception noted in \S\ref{sec:build-theory}.  All dependencies are resolved statically before
execution, placing OxyMake in the same scheduler column as Make, Ninja
and Buck rather than the restarting column that holds Bazel or the
suspending column occupied by Shake.  Unlike
Make, OxyMake replaces dirty-bit rebuilding (re-run whenever any input
timestamp changes) with content-addressed verifying traces, and unlike
Bazel, it targets scientific workflows with declarative TOML rules,
wildcard expansion, and environment management.

\subsection{Intended Use}

The design fits teams that already express computation as deterministic
commands with files as artefact boundaries and want to avoid recomputation
without deploying a service. Two grounds are the most credible first
targets. Portable scientific pipelines meet a mechanical consequence of
the scope of a per-tree provenance record (\S\ref{sec:introduction}),
which we state rather than survey: such a record does not survive a
fresh checkout or a move to a second machine. HPC deployments on shared filesystems are \emph{not} a
first target in v0.1.0: the local-disk requirement on \code{.oxymake/}
(\S\ref{sec:limitations}) excludes them until the state directory is
relocatable. Small
data teams running a modest batch DAG face a deployment cost in the
service-oriented alternatives, which are built around a persistent
scheduler and its supporting services~\cite{airflow}; that
infrastructure is disproportionate to a single batch job. ML
feature-and-sweep pipelines and agent-driven runs are plausible fits by
the same reasoning, but are not demonstrated here: OxyMake integrates
alongside experiment trackers rather than replacing them, and the MCP
interface is a driving surface, not a durability layer
(\S\ref{sec:agent-api}). These are design-fit inferences from the
engine's contract, not adoption results.

\subsection{Limitations and Future Work}
\label{sec:limitations}

\paragraph{Gate enforcement.}
Gates are declared in the Oxymakefile, listed and decided with
\code{ox gate}, and the scheduler carries the logic to hold a guarded
job until its gate is approved; but in v0.1.0 no shipped command passes a
gate checker to the scheduler, so a guarded rule runs without approval
and no pending gate is ever registered (\S\ref{sec:agent-api}). An
operator who declares a gate as a network or overwrite checkpoint gets
neither protection nor a signal that the gate is inert. The fix is a
checker backed by the state database, keyed by gate name rather than by
the numeric record identifier the current table uses, plus a \code{lint}
warning while enforcement is unwired (issue~\#2).

\paragraph{TOML expressiveness.}
The decision to use TOML rather than a Turing-complete DSL limits the
expressiveness of workflow definitions. Complex configuration generation
must happen outside the Oxymakefile via scripts. While this preserves static
parseability, it may frustrate users accustomed to Snakemake's Python
flexibility. We note that the choice is not strictly binary: languages such
as Starlark~\cite{starlark} (deterministic, sandboxed Python subset used by
Bazel), CUE~\cite{cue} (constraint-based configuration with types and
validation), and Dhall~\cite{dhall} (a total functional language for
configuration) occupy a middle ground between inert data formats and
general-purpose languages, offering controlled expressiveness with static
analysis guarantees. OxyMake opts for the simplest end of this spectrum---plain
TOML---because workflow \emph{specification} rarely needs computation; when it
does, a config generation pattern (\code{python gen\_config.py > config.toml})
or a minimal expression language (pure functions, no loops) provides an escape
hatch without compromising the static parseability of the Oxymakefile itself.

\paragraph{No sandbox: undeclared inputs are trusted.}
As stated in the threat model (\S\ref{sec:undeclared-inputs}), OxyMake
does not isolate rule execution, so the cache key's completeness is
only as good as the user's input declaration: an undeclared read is a
silent stale-reuse hazard. An opt-in sandboxed execution mode
(namespace or seatbelt isolation that turns an undeclared read into a
build-time error, recovering Nix's enforcement property) is future work;
until it lands, the mitigation is declaration discipline plus the
auditable \code{ox.lock} record.

\paragraph{SQLite on network filesystems.}
SQLite WAL mode does not work on NFS, Lustre, or GPFS---precisely the
filesystems used on HPC clusters where Slurm runs. OxyMake requires
\code{.oxymake/} to reside on local disk; the scheduler runs on the
submission node, and compute nodes never touch SQLite. The requirement
is currently unassisted: \code{.oxymake/} is created in the working
directory, and no flag, config key, or environment variable relocates
it, so the state database cannot be separated from the workspace whose
inputs and outputs are resolved against the same directory. An operator
who needs the project tree on shared storage has no supported way to
keep only the database local; a relocatable state directory is future
work alongside multi-node coordination (multiple submission nodes),
which requires a future coordination layer. This local-disk requirement
is also the discharge of the \code{StateDbAtomicCommit} axiom that the
model-checked multi-session guarantee rests on
(\S\ref{sec:named-invariants}).

\paragraph{Cache-key residual exclusions.}
Two ingredients are deliberately excluded from the cache key
(Eq.~\ref{eq:cache-key}). One is \code{call}-mode function bodies: the
key covers the function \emph{reference} (module path and name), but the
module's source file is content-tracked only when declared as an input.
Editing an imported module without redeclaring it can therefore serve a
stale cached result; shell, inline-\code{run}, and script modes have no
such gap (script file content is hashed into the key). The other is
mutable container image tags: a \code{docker} or \code{apptainer} environment is
hashed as the literal image reference, without resolving tags to
digests---resolving would require a container-runtime round-trip per key
computation and would break offline runs. A re-pushed
\code{python:3.12-slim} therefore does not invalidate the cache; users
who need this guarantee should pin images by digest
(\code{python@sha256:\ldots}), which the key then captures exactly.

\paragraph{In-memory passing limitations.}
The \code{call} mode's in-memory passing supports only types
representable in the configured serialization format. Arbitrary Python
objects (e.g., trained ML models) must be serialized to disk.
A \code{pickle} codec provides a fallback but sacrifices cross-language
compatibility.

\paragraph{Distributed executors.}
The local executor is exercised throughout this evaluation. The Slurm
and Ray executors are implemented but have no supported deployment
configuration in v0.1.0, because both need the workspace visible to the
compute nodes while \code{.oxymake/} must be on local disk
(\S\ref{sec:distributed-related}); a relocatable state directory is the
prerequisite. The Kubernetes
executor is designed (Section~\ref{sec:plugins}) but not yet implemented.
The \code{Executor} trait and capability negotiation are in place; what
remains is the backend-specific job submission, status polling, and log
retrieval code for the Kubernetes backend.

\paragraph{Optimization passes.}
Only cache pruning is implemented; the five further passes of
Section~\ref{sec:three-graphs} are designed but await implementation.
The current scheduler dispatches jobs in topological order without
them; they would reduce execution time for \code{call}-mode workflows
but do not affect correctness.

\paragraph{Evaluation statistics.}
The evaluation is limited in statistical depth as well as in scope.
All measurements come from a single machine (Apple M4~Max, macOS,
arm64); a Linux/x86-64 replication has not been performed, and no
cross-platform claim is made. Each end-to-end time is a single timed
run and resolution times are \code{hyperfine} medians; dispersion
(IQR or confidence intervals) is not reported. The benchmark's
rendered summary prints the 10{,}001-job end-to-end times rounded to
minutes scale; Table~\ref{tab:e2e} quotes the same samples in seconds.
Only the Snakemake~7.x line was measured; a run against the current 9.x
line is future work. The reported ratios
should be read with these limits in mind.

\paragraph{Maturity.}
OxyMake is in active development. The core engine
(Table~\ref{tab:crate-size}) implements the complete DAG pipeline from
TOML parsing through parallel scheduling and execution
(Section~\ref{sec:evaluation}). The evaluated
implementation is pre-1.0; its API may change.

\subsection{Future Work: Substrate Composition}
\label{sec:future-work}

Future work will study whether OxyMake's orchestration-level cache
contract composes with input-addressed (functional) package stores such
as Guix~\cite{courtesGuixHPCReproducible2015,dolstraPurelyFunctionalSoftware2006},
whose store paths hash all inputs of the build that produced them.
OxyMake's cache key records an \emph{output-equivalence} relation
(\S\ref{sec:content-addressable}), while such a substrate
records an \emph{input-equivalence} relation: does this input hash
produce this binary? The conjecture is that the two witnesses are
orthogonal---each checkable independently, neither subsuming the
other---so that their conjunction is a strictly stronger
reproducibility statement than either alone, of the kind the guix-cwl
reference workflows~\cite{prinsGuixCWLPipelines2018} illustrate at the
substrate layer. This hypothesis is not evaluated here.

Rolling a substrate into the engine is explicitly \emph{not} the
intended path. Each layer of the reproducibility ladder
(Table~\ref{tab:fair-ladder}) has a distinct owner in the broader
ecosystem---Guix-HPC for L1, build-systems literature for L3, the
RO-Crate community~\cite{soilandReyesPackagingResearch2022} for
L4---and an engine that collapses all four into one binary inherits
all four maintenance burdens. A Guix-backed execution environment and
a CWL reader/writer for \code{ox translate} (which today ships
Snakemake and WDL bridges only) are therefore framed as invitations to
a community-driven open-source effort rather than commitments of this
paper.

\subsection{A Note on Polyglot Layering}
\label{sec:language-material}

OxyMake assigns each layer a language chosen for the layer's
constraints, not for uniformity. The orchestration core is Rust:
the type system encodes pipeline-state invariants
(\code{Rule} vs \code{ConcreteJob}) at compile time, and \code{Send +
Sync} bounds rule out data races on the scheduler's shared state at
compile time (\S\ref{sec:rust}). The workflow specification is TOML: deliberately not
Turing-complete, statically parseable in microseconds, and analysable
without execution---a property Snakemake's Python DSL cannot offer.
Rule bodies execute in Python, R, Julia, or shell, selected per rule
for the domain libraries each ecosystem encodes.

Each subsystem uses the language whose properties match its
function, with serialisation contracts (structured IPC, file I/O) at
the boundaries. The cost is an extra language frontier per layer; the
benefit is that no single language has to compromise.

\section{Conclusion}
\label{sec:conclusion}

We have presented OxyMake, a workflow engine built on the position that
the validity of a result is a property of the work rather than of the
place it ran, and that this need not cost an operational system. The
computation key is a function of declared content rather than a per-tree
record: when the entire declared specification---rule source, inputs,
parameters, environment, shell, platform---is unchanged, the key is
identical in any checkout under the same key-format version. Identity
is not reuse; the rebuild decision a checkout takes also consults its
materialised outputs, the selected validation policy, and---under the
\code{mtime+hash} default and \code{hash}---its local key-to-output
index, which the opt-in \code{mtime} policy does not read
(\S\ref{sec:content-addressable}). The workflow is
a declarative TOML file
executed by a single 17.4\,MB Rust binary with no daemon, and every
command can emit structured NDJSON with \code{--json} for a human or a
program. \code{ox run} ensures the declared outputs exist, runs only
the jobs that are missing or out of date, and is safe to repeat.

The measured costs and benefits are bounded. On a
synthetic 10,001-job workflow on one machine, DAG resolution takes 118\,ms
against Snakemake~7.32.4's 2.72\,s; a cold end-to-end run is
1.47--2.57$\times$ slower, and a warm no-op re-run is 2.52$\times$
faster under the shipped \code{mtime+hash} default, 6.95$\times$ under
\code{mtime} and 1.91$\times$ under \code{hash}. Three TLA+ specifications
model-check bounded safety of the concurrent state protocol. The cache
key is computed from the declared specification only, so an undeclared
read can cause stale reuse; output re-verification depth is a policy;
and the claim protocol is not yet the scheduling gate. These
limits are stated where they apply (\S\ref{sec:limitations}).

OxyMake is open-source software under the Apache-2.0/MIT dual license,
available at \code{https://github.com/noogram/oxymake}
(\code{https://oxymake.noogram.dev}).

\section*{Acknowledgments}

We thank the GNU Guix, CWL, and FAIR-workflows communities, whose prior
work informed this engine's design.

We thank Michael R. Crusoe for comments on the descriptions of CWL and
\code{cwltool}.

\section*{Disclosure of AI assistance}

The development of OxyMake and the preparation of this manuscript made
substantial use of large language model assistants, for code, prose
drafting, and successive rounds of adversarial review. All design
decisions, measurements, and final wording were verified by the author,
who takes full responsibility for the content.

\appendix

\section{Crate Census}
\label{app:crates}

\begin{table}[htbp]
\centering\small
\begin{tabular}{@{}lrrr@{}}
\toprule
\textbf{Crate} & \textbf{Lines} & \textbf{Tests} & \textbf{Doc Tests} \\
\midrule
\code{ox-core}         & 15{,}568 & 550 & 29 \\
\code{ox-cli}          & 11{,}392 & 211 &  0 \\
\code{ox-translate}    &  6{,}170 & 162 &  2 \\
\code{ox-exec-ray}     &  5{,}383 & 156 &  1 \\
\code{ox-exec-local}   &  4{,}602 & 103 &  1 \\
\code{ox-state}        &  3{,}862 &  70 & 10 \\
\code{ox-format}       &  3{,}497 & 145 &  0 \\
\code{ox-exec-slurm}   &  3{,}382 &  92 &  1 \\
\code{ox-cache}        &  2{,}279 &  99 &  0 \\
\code{ox-mcp}          &  1{,}901 &  85 &  0 \\
\code{ox-dashboard}    &  1{,}154 &  32 &  1 \\
\code{ox-monitor-tui}  &  1{,}049 &  23 &  3 \\
\code{ox-report-term}  &     926 &  30 &  3 \\
\code{ox-plan}         &     681 &  25 &  6 \\
\code{ox-lock}         &     680 &  11 &  0 \\
\code{ox-api}          &     473 &  10 &  2 \\
\code{ox-cache-remote} &     455 &  15 &  0 \\
\code{ox-report-json}  &     389 &  20 &  1 \\
\code{ox-metrics}$^\dagger$ &     277 &   8 &  3 \\
\code{ox-codec-core}   &     275 &  20 &  0 \\
\code{ox-render}       &     176 &   6 &  0 \\
Stub crates (\code{ox-env-*}, \code{ox-storage-local}) & 25 & 3 & 0 \\
\code{oxymake} (name reservation) & 0 & 0 & 0 \\
\midrule
\textbf{Total}         & \textbf{\metricSLOC} & \textbf{1{,}876} & \textbf{\metricDocTests} \\
\bottomrule
\end{tabular}
\caption{Crate architecture with implementation size and test counts.
  Lines are Rust code lines (\code{tokei}); Tests are passing unit and
  integration tests per crate (\code{cargo test}); doc tests are listed
  separately. Columns sum to the Total row.
  $^\dagger$\code{ox-metrics} is excluded from the Cargo workspace for
  dependency isolation and is tested via its own manifest; the
  workspace test run therefore reports \metricTests{} tests (the table
  totals minus \code{ox-metrics}'s 8 tests and 3 doc tests).}
\label{tab:crate-size}
\end{table}

\begin{figure}[htbp]
  \centering
  \includegraphics[width=0.8\textwidth]{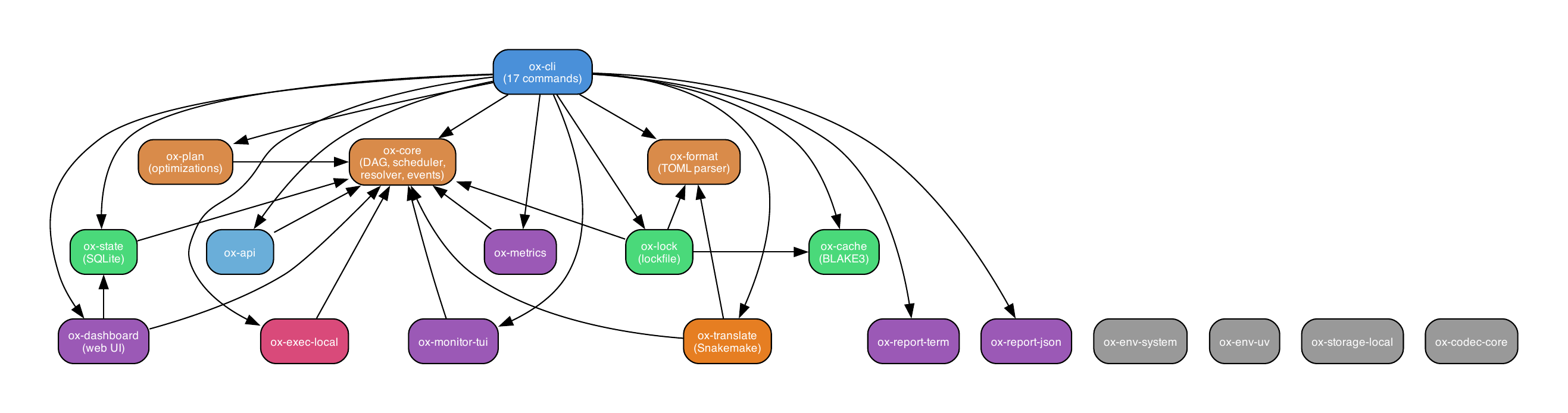}
  \caption{Dependency graph of the principal crates in OxyMake's workspace.
    \code{ox-cli} sits at the top as the entry point, delegating to
    \code{ox-core} (engine), \code{ox-format} (TOML parser),
    \code{ox-state} (SQLite persistence), and specialized crates for
    execution, reporting, monitoring, and translation.}
  \label{fig:crate-arch}
\end{figure}

\noindent The size distribution reveals the expected concentration:
\code{ox-\allowbreak{}core} contains 24\% of the codebase (DAG algorithms,
scheduler, type system, event bus) and 29\% of the unit and integration
tests. Three stub
crates (\code{ox-\allowbreak{}env-\allowbreak{}system},
\code{ox-\allowbreak{}env-\allowbreak{}uv}, \code{ox-\allowbreak{}storage-\allowbreak{}local})
contain trait scaffolding only; the functionality they will host is
implemented inline today (\S\ref{sec:plugins}).
Figure~\ref{fig:crate-arch} illustrates the dependency relationships
between crates.

\section{TLA+ Specification Inventory}
\label{app:reproducibility}

Table~\ref{tab:tla-specs} summarizes the model-checking evidence
behind \S\ref{sec:named-invariants}: for each specification, the
safety properties it checks, the bounds at which TLC explored the
state space exhaustively, and the size of that exploration. The
specifications, their TLC configuration files, and the archived TLC
outputs are in the public repository under \code{spec/tla/}.

\begin{table}[htbp]
\centering\small\setlength{\tabcolsep}{4pt}
\begin{tabular}{@{}lp{0.24\linewidth}p{0.20\linewidth}rr@{}}
\toprule
\textbf{Module} & \textbf{Properties} & \textbf{Bounds} & \textbf{States} & \textbf{Depth} \\
\midrule
\code{Cache\allowbreak Consistency.tla}  & OX-1, OX-6 (key determinism, stationary cache safety) & 2 workers $\times$ 3 rules, worker crashes & 7\,436 & 16 \\
\addlinespace
\code{Cooperative\allowbreak Claim.tla}  & INV-2 (claim atomicity, stale-session reclaim, zombie terminal writes) & 3 sessions $\times$ 2 jobs, TTL 2, clock bound 4 & 1\,663\,056 & 23 \\
\addlinespace
\code{Cancel\allowbreak Propagation.tla} & INV-3 (\code{Cancelled\allowbreak Never\allowbreak Cached},
  \code{Job\allowbreak Failed\allowbreak Implies\allowbreak No\allowbreak Intent},
  \code{No\allowbreak Zombie\allowbreak Running}) & 3 jobs & 32\,768 & 22 \\
\bottomrule
\end{tabular}
\caption{TLA+ specifications and their model-checking scope. States =
  distinct states explored by TLC at the stated bounds; Depth = BFS
  depth of the complete state graph. All checks assume the substrate
  axioms in the right column of Table~\ref{tab:verified-vs-assumed}
  (SQLite commit atomicity on a local filesystem, executor honesty,
  atomic storage deletion, key purity, and input-declaration
  completeness). Two falsification configurations accompany the suite:
  each models a named defect (a terminal \code{UPDATE} without session
  filter; an undeclared input leaking into the cache key) and is
  expected to fail, which shows that the corresponding passing
  specification checks a property the defect can violate.}
\label{tab:tla-specs}
\end{table}

\clearpage
\bibliographystyle{plain}
{\hbadness=10000          
\bibliography{references}
}

\end{document}